\documentclass[amsmath,nofootinbib,amssymb,prd,superscriptaddress]{revtex4}
\pdfoutput=1

\usepackage[utf8]{inputenc}
\usepackage{float}
\usepackage{epsfig}
\usepackage{graphicx}
\usepackage{color}
\usepackage{tabularx}
\usepackage{color}
\usepackage{verbatim}
\usepackage[inline, final]{showlabels}
\usepackage{tensor}
\usepackage{enumerate}

\usepackage{subcaption}
\usepackage{color}
\usepackage[bookmarks,colorlinks,linkcolor=webblue,citecolor=webgreen]{hyperref}	
\definecolor{webgreen}{rgb}{0, 0.5, 0} 
\definecolor{webblue}{rgb}{0, 0, 0.5} 
\definecolor{webred}{rgb}{0.5, 0, 0} 

\newcommand{\av}[1]{\left\langle #1 \right\rangle}

\newcommand{\ee}{\text{e}} 

\newcommand{\di}{\text{d}}

\newcommand{\pd}[2]{\frac{\partial #1}{\partial #2}}

\makeatletter
\g@addto@macro\bfseries{\boldmath}  
\makeatother

\renewcommand{\v}[1]{\boldsymbol{#1}}

\renewcommand{\d}[2]{\frac{\text{d} #1}{\text{d} #2}}
\renewcommand{\pd}[2]{\frac{\partial #1}{\partial #2}}

\begin{document}
\preprint{}
\title{Can late-time extensions solve the $H_0$ and $\sigma_8$ tensions?}
\author{Lavinia Heisenberg\footnote{lavinia.heisenberg@phys.ethz.ch}}
\affiliation{Institut f\"{u}r Theoretische Physik, Philosophenweg 16, 69120 Heidelberg, Germany}
\affiliation{Institute for Theoretical Physics, ETH Z\"{u}rich, Wolfgang-Pauli-Strasse 27, 8093, Z\"{u}rich, Switzerland}
\author{Hector Villarrubia-Rojo\footnote{herojo@phys.ethz.ch}}
\author{Jann Zosso\footnote{jzosso@phys.ethz.ch}}
\affiliation{Institute for Theoretical Physics, ETH Z\"{u}rich, Wolfgang-Pauli-Strasse 27, 8093, Z\"{u}rich, Switzerland}
\date{\today}

\begin{abstract}
We analyze the properties that any late-time modification of the $\Lambda$CDM expansion history 
must have in order to consistently solve both the $H_0$ and the $\sigma_8$ tensions. 
Taking a model-independent approach, we obtain a set of necessary conditions that can be applied to generic
late-time extensions.
Our results are fully analytical and merely based on the assumptions that the deviations 
from the $\Lambda$CDM background remain small. For the concrete case of a dark energy fluid with equation of state $w(z)$, we derive the following general 
requirements: (i) Solving the $H_0$ tension demands $w(z)<-1$ at some $z$ (ii) Solving both 
the $H_0$ and $\sigma_8$ tensions requires $w(z)$ to cross the phantom divide. 
Finally, we also allow for small deviations on the effective gravitational constant.
In this case, our method is still able to constrain the functional form 
of these deviations.
\end{abstract}

\maketitle
\tableofcontents

\section{Introduction}\label{sec:intro}

Historically, the quest for a satisfactory description of our universe has always been 
guided by the latest observational data. It is therefore not surprising that the dramatic 
increase of the quantity and quality of cosmological observations over the last 25 years 
has allowed for a revolution on the theoretical side as well. The $\Lambda$CDM model has 
emerged as the leading theoretical description of cosmic evolution, explaining key features,
such as the distribution of the cosmic microwave background (CMB) anisotropies, with only a 
few free parameters. However, despite its success, several observations have been, and still 
are, hard to account for within this paradigm. In particular, the well-known $H_0$ tension, 
that is the discrepancy of the Hubble constant inferred from the CMB within $\Lambda$CDM 
\cite{Planck:2018vyg} compared to the results of local measurements \cite{Riess:2019cxk,
Pesce:2020xfe,Wong:2019kwg, Riess:2020fzl}, has become increasingly worrying in recent years and it is 
hard to disregard it as a simple statistical fluke. At this point, either there is 
something wrong with different, independent observations or we must change the theoretical 
framework to interpret them.
	
Another mayor concern in the community is the $\sigma_8$ tension which, just as the $H_0$ 
tension, arises when comparing the CMB-inferred value of the clustering amplitude to 
alternative observations, in this case large scale structure (LSS) surveys 
\cite{DES:2017myr, DES:2021wwk, KiDS:2020suj, Heymans:2020gsg}. Focusing on these two parameters and disregarding 
correlations with other parameters, we may say that CMB data favors a lower value of $H_0$ 
while at the same time preferring a higher $\sigma_8$ value compared to late-time 
measurements. See \cite{Nunes:2021ipq} for a compilation of recent measurements.
	
In recent years, great efforts have been made towards solving the $H_0$ tension, see
e.g. \cite{Poulin:2018cxd, Smith:2019ihp, Alcaniz:2019kah, Zumalacarregui:2020cjh, 
Gomez-Valent:2020mqn, Ballesteros:2020sik, Jimenez:2020bgw, DiValentino:2020naf,
Banerjee:2020xcn, Krishnan:2021dyb, Teng:2021cvy, Ballardini:2021evv,
Braglia:2020auw, Braglia:2020iik}, 
as well as the $\sigma_8$ tension, see e.g. \cite{Lambiase:2018ows, 
Keeley:2019esp, DiValentino:2019ffd, Jedamzik:2020zmd, Clark:2021hlo, SolaPeracaula:2021gxi,
Alestas:2021xes, Nunes:2021ipq, Schoneberg:2021qvd, Alestas:2021luu, Ye:2021iwa}
(see also \cite{Riess:2019qba, Knox:2019rjx, DiValentino:2020vvd, DiValentino:2020zio,
DiValentino:2021izs, Perivolaropoulos:2021jda} for an overview of observations and models). 
Among these solutions, those that modify the behaviour of dark energy either at late or 
early times have received special attention. Both run into similar problems when
trying to solve the Hubble tension while being consistent with complementary observations,
and in this work we will restrict ourselves to late-time models. 
Typical late-time solutions to the $H_0$ tension, as for example models based on scalar or vector Galileons 
\cite{Renk:2017rzu,Frusciante:2019puu,deFelice:2017paw,DeFelice:2020sdq,Heisenberg:2020xak}, 
usually lead to an even larger value of $\sigma_8$ than within $\Lambda$CDM and 
therefore potentially increase the $\sigma_8$ tension. One should however keep in mind that 
the $\sigma_8$ tension is only properly addressed as a tension in the $\sigma_8$-$\Omega_m$ 
plane, or in the general multidimensional posterior. While these models predict a 
slightly larger $\sigma_8$ they also prefer smaller values for $\Omega_m$, a result in the 
line of weak-lensing surveys. However, even if one could argue that strictly 
speaking such late-time solutions to the Hubble tension are still statistically compatible 
with current $\sigma_8$ measurements, they clearly show the wrong trend, such that they 
will most likely be difficult to reconcile with a low $\sigma_8$ value if it is
confirmed by the more precise measurements of the next generation of LSS surveys.
	
This generic trend, that late-time dark energy models easing the Hubble tension 
predominantly increase $\sigma_8$ as well, can be understood as follows. A realistic dark 
energy model typically affects $\sigma_8$ in two ways: 1) Through its effects on the 
expansion history, i.e. modifications of the background equation of state. 2) Through its 
clustering properties, i.e. clustering dark energy that can modify the effective Newton 
constant $G_\text{eff}$ which governs the evolution of the matter growth function. In the 
models mentioned above with a phantom equation of state both effects contribute to an 
increase in $\sigma_8$: 1) A phantom-like evolution of dark energy extends the 
matter-dominated phase, boosting the matter growth. 2) Dark energy clusters at late times, 
increasing $G_\text{eff}$ and further boosting the amplitude of perturbations. The key 
point is that the same phantom-like equation of state that is crucial to solve the $H_0$ 
tension, can only worsen the $\sigma_8$ tension.
	
In face of these problems, one may wonder if it is even possible to solve both tensions 
modifying only the late-time dark energy behavior. Of course, in a consistent dark energy 
model one should not only study the background evolution, but the perturbations as well. 
And while the background evolution is governed by the dark energy equation of state $w(z)$, 
the perturbations are also affected by the dark energy sound speed $c_\text{s}(z)$. Hence, 
at first sight one could conclude that with two arbitrary functions at hand it should not be 
difficult to find a dark energy model that solves both tensions at once. However, this is 
not the case since in realistic scenarios such as vector Galileons both functions are not 
independent and their observational impact is very different. In fact, while $c_\text{s}(z)$ 
is relevant for observables like the ISW effect, the modifications in $w(z)$ are the main 
force driving the values of $\sigma_8$ and $H_0$. In the light of these considerations we 
will therefore narrow down the scope of this work by mostly neglecting the effects of dark 
energy perturbations and address the following main question:\\
	
\noindent\emph{Can the $H_0$ and $\sigma_8$ tensions be simultaneously relieved modifying only the 
dark energy equation of state $w(z)$ at late times?}\\
	
We will show that the answer is no if the dark energy equation of state does not meet some 
very definite criteria. These conclusions apply to any dark energy model in which the 
perturbations do not play a leading role in the determination of $\sigma_8$. After this,
we will generalize the results to the case where dark energy also affects the growth
of structure through a change in the effective gravitational constant. Thus, our 
results provide valuable insights into the behavior of the dark sector and can be seen as 
hints towards building successful models beyond $\Lambda$CDM.\\
	
The main steps of the computation can be succintly summarized as follows:
	\begin{enumerate}[i)]
		\item We will start off with a late-time $\Lambda$CDM cosmology, that can
			effectively be described by two free parameters $\big(h, \omega_m\big)$, i.e.
			the Hubble constant and the matter abundance. In a second step, we consider an 
			alternative cosmology with slightly different parameters and with a different
			expansion history $\big(h+\delta h, \omega_m+\delta\omega_m, \delta H(z)\big)$.
			Here, $\delta H(z)$ is an \emph{arbitrary} function that produces a small 
			deformation	of the $\Lambda$CDM expansion history, for fixed $h$ and $\omega_m$. 
			Restricting ourselves to late time modifications translates into the assumption 
			that $\delta H(z) = 0$ for roughly $z>300$.
			With all the deformations considered to be small, the Hubble parameter
			in the alternative cosmology can be written as 
			\begin{equation}
				H = H_\text{\tiny $\Lambda$CDM} + \Delta H\big(\delta h, \delta\omega_m, \delta H(z)\big)\ .
			\end{equation}
		\item Working to first order, we compute the variations induced by the modified
			Hubble parameter in different cosmological observables.
		\item Because of the deformation $\delta H(z)$, the observationally preferred 
			values for $h$ and $\omega_m$ in the new cosmology will be different compared to the initial $\Lambda$CDM model. 	
			The variations $\delta h$ and $\delta\omega_m$ can be related to $\delta H(z)$ by choosing two very well 							measured observables whose value should not change in the new cosmology, i.e. we impose
			their variation to be zero in order to be compatible with observations. This allows to compute the
			response functions
			\begin{subequations}
			\begin{align}
				\frac{\delta h}{h} &= \int\mathcal{R}_h(z)\frac{\delta H(z)}{H(z)}\frac{\di z}{1+z}\ ,\\
				\frac{\delta \omega_m}{\omega_m} &= \int\mathcal{R}_{\omega_m}(z)\frac{\delta H(z)}{H(z)}\frac{\di z}{1+z}\ .
			\end{align}
			\end{subequations}
			In this work we will choose the variations of the CMB distance priors 
			\cite{Chen:2018dbv} to vanish. The response functions $\mathcal{R}_h$ and
			$\mathcal{R}_{\omega_m}$ are fully analytical and are defined in 
			\eqref{eq:def_Rh_Rom}.
		\item Based on the results above we can then compute the response function of any other 
			quantity, and crucially in our case 
			\begin{equation}
				\frac{\Delta\sigma_8}{\sigma_8} = \int\mathcal{R}_{\sigma_8}(z)\frac{\delta H(z)}{H(z)}\frac{\di z}{1+z}\ .
			\end{equation}
			Depending on the shape of the response functions, this allows to derive general 
			requirements on the functional form of $\delta H(z)$ in order to achieve 
			the desired variations in $h$ and $\sigma_8$. Again, $\mathcal{R}_{\sigma_8}$ can
			be computed analytically and it will be derived in Section \ref{sec:sigma8}.
			
		\item In Section \ref{sec:G_eff}, we will generalize these results and include a second free 
			function, $\delta G_\text{eff}(z)$, that affects the evolution of 
			$\sigma_8$, computing its associated response function along the same lines.
	\end{enumerate}	
	
	This paper is organized as follows. In Section \ref{sec:variations} we introduce
	the deformations of the background and most of the notation. Section \ref{sec:CMBpriors}
	will cover the choice of observational data (CMB priors) used to compute
	the response functions. Section \ref{sec:sigma8} deals with the variations of the
	growth factor and $\sigma_8$. It also includes a generalization for	models that modify 
	the effective Newton constant.  In Section \ref{sec:supernovae} we address the resolution of the $H_0$ tension in more detail, 			analyzing the differences that arise
	when formulated in terms of the supernova absolute magnitude $M$. Section 
	\ref{sec:summary} summarizes the results of this work. Appendix \ref{sec:app_formulae}
	collects all the analytical formulae used to derive the results in the main text.
	Finally, in Appendix \ref{sec:app_benchmarks} we present some tests 
	performed to check the accuracy of the first-order, analytical results against
	the full numerical computation with a Boltzmann code for a particular dark 
	energy model.
	
	
\section{Deformations of the expansion history}\label{sec:variations}
	The Hubble parameter in $\Lambda$CDM can be written as
	\begin{align}
		H^2_\text{\tiny $\Lambda$CDM} &= H_0^2 \Big(\Omega_m (1+z)^{3} + \Omega_r (1+z)^{4} + \Omega_\Lambda\Big)\nonumber\\
			&= C_H^2 \Big(\omega_m (1+z)^{3} + \omega_r (1+z)^{4} + \omega_\Lambda\Big)\ ,
	\end{align}	
	where $C_H\equiv 100\ \text{km}\,\text{s}^{-1}\,\text{Mpc}^{-1}$ and 
	\begin{equation}
		\omega_\Lambda = h^2-\omega_m-\omega_r\ .
	\end{equation}
	Let us consider now a generic extension that slightly modifies the expansion history,
	for fixed values of all the cosmological parameters, so the new Hubble parameter is
	\begin{equation}
		H(h, \omega_m) = H_\text{\tiny $\Lambda$CDM}(h, \omega_m) + \delta H\ .
	\end{equation}
	This deformation of the expansion history will also shift the preferred values for
	the $\Lambda$CDM parameters, by a small amount. The observationally preferred 
	background in $\Lambda$CDM and in the generic extension can then be related as
	\begin{equation}
		H(h+\delta h, \omega_m +\delta \omega_m) = H_\text{\tiny $\Lambda$CDM}(h, \omega_m)
			+ \Delta H\ ,
	\end{equation}
	where we are also assuming that $\delta H$ only produces late-time changes so the 
	cosmology can be effectively described by $h$ and $\omega_m$.
	Assuming that all these variations are small and working to first order we have
	\begin{equation}\label{eq:DeltaH}
		\frac{\Delta H}{H} = \frac{H_0^2}{H^2}\frac{\delta h}{h} 
								+ m(z)\frac{\delta\omega_m}{\omega_m}
								+ \frac{\delta H}{H}\ ,\qquad\quad
		m(z)\equiv \frac{\Omega_m H_0^2}{2H^2}\left((1+z)^{3}-1\right)\ .
	\end{equation}
	
	Starting with the general variation \eqref{eq:DeltaH}, we can propagate its
	effect to any cosmological observable. In general, for every cosmological quantity $g(z)$ we will express its
	variation as
	\begin{equation}
		\frac{\Delta g(z)}{g(z)} = I_g(z)\frac{\delta h}{h}
			+ J_g(z)\frac{\delta\omega_m}{\omega_m}
			+ \int^\infty_0\frac{\di x_z}{1+x_z}R_g(x_z, z)\frac{\delta H(x_z)}{H(x_z)}\ .
	\end{equation}
	For instance, from the definition of the conformal, luminosity and angular diameter
	distances,
	\begin{subequations}
	\begin{align}
		\chi(z) &=\int^z_0\frac{\di z}{H(z)}\ ,\\
		d_L(z)  &= (1+z)\chi(z)\ ,\\
		d_A(z)  &= \frac{1}{1+z}\chi(z)\ ,
	\end{align}
	\end{subequations}
	we can easily compute
	\begin{equation}
		\left\{\begin{array}{l}
		\displaystyle
			I_{\chi}(z) = I_{d_L}(z) = I_{d_A}(z) = -\frac{1}{\chi(z)}\int^z_0\di x_z\frac{H_0^2}{H^3}\ ,\\[8pt]
			\displaystyle
			J_{\chi}(z) = J_{d_L}(z) = J_{d_A}(z) = -\frac{1}{\chi(z)}\int^z_0\di x_z\frac{H_0^2}{H^3}m(x_z)\ ,\\[8pt]
			\displaystyle
			R_{\chi}(x_z,z) = R_{d_L}(x_z,z) = R_{d_A}(x_z,z) = -(1+x_z)\frac{\theta(z-x_z)}{\chi(z)H(x_z)}\ .
		\end{array}\right.
	\end{equation}
	Notice that, since we are working to first order, every function like $\chi(z)$ and
	$H(z)$ can be computed in the base $\Lambda$CDM cosmology. The full analytical 
	expressions for all the $(I, J, R)$ functions that are used in this work are
	collected in Appendix \ref{sec:app_formulae}. 

	Finally, notice that we can also express the previous results in terms of 
	a variation on the energy content
	\begin{equation}
		H^2(h, \omega_m) = H^2_{\text{\tiny $\Lambda$CDM}}(h, \omega_m) + H_0^2\delta\Omega \ .
	\end{equation}
	Working again to first order we have
	\begin{equation}
		\frac{\delta H}{H} = \frac{H_0^2}{2H^2}\delta\Omega\ .
	\end{equation}
	Note that we did not, and will not, specify a functional form for either
	$\delta H$ or $\delta\Omega$. 
	However, in particular models some additional properties may be desirable. For
	instance, if $\delta\Omega$ arises from a dark energy model, we may want to 
	require that the dark energy density is positive, leading to 
	\begin{equation}
		\Omega_\text{DE}(z)\equiv \Omega_\Lambda + \delta\Omega(z) > 0\qquad \text{(DE model)}
	\end{equation}
	In this case, we can also relate the variation to the equation of state of dark energy
	$w(z)$
	\begin{equation}\label{eq:Omegaz_w}
		\delta \Omega(z) = \Omega_\Lambda \left\{\exp\left(3\int^z_0\big(1+w(z)\big)\frac{\di z}{1+z}\right)-1\right\}\ .
	\end{equation}
	Following the same reasoning, in a model with dark matter and dark energy interactions we have
	\begin{equation}
		\Omega_\text{DE-DM}(z)\equiv \Omega_\Lambda +\Omega_\text{cdm}(1+z)^3 + \delta\Omega(z) > 0\qquad 
			\text{(Interacting DM-DE model)}
	\end{equation}

\section{CMB priors and the $H_0$ tension}\label{sec:CMBpriors}
	In the previous section we considered a generic background modification over
	a $\Lambda$CDM cosmology. However, we know that the extremely precise observations
	of the CMB severely restrict such modifications. Two combinations of parameters
	are particularly well measured, 
	\begin{subequations}
	\begin{align}
		\theta_* &\equiv \frac{r_\text{s}(z_*)}{(1+z_*)d_A(z_*)}\ ,\\
		R_* &\equiv (1+z_*)d_A(z_*)\sqrt{\Omega_m H_0^2}\ .
	\end{align}
	\end{subequations}
	where $r_s(z)$ is the comoving sound horizon
	\begin{equation}
		r_\text{s}(z) =\int^\infty_z\frac{\di z}{H}c_\text{s}\ ,\qquad c_\text{s}=\frac{1}{\sqrt{3(1+R)}}\ ,\qquad
		R=\frac{3\Omega_b}{4\Omega_\gamma(1+z)}\ ,
	\end{equation}
	and $z_*\simeq 1090$ is the redshift at decoupling, see \cite{Chen:2018dbv} for a more accurate
	interpolation formula.
	These are commonly referred to as the CMB distance priors: the acoustic scale ($\theta_*$) and the shift 
	parameter ($R_*$), 
	that govern the angular position and the relative heights of the peaks, respectively.
	Their latest values using the Planck 2018 release, for $\Lambda$CDM and some
	extensions, can be found in \cite{Chen:2018dbv}.
	We can compute their variation following the steps of the previous section
	\begin{subequations}\label{eq:CMBpriors_variation}
	\begin{align}
		\frac{\Delta \theta_*}{\theta_*} &= \frac{\Delta r_\text{s}^*}{r_\text{s}^*} - \frac{\Delta d_A^*}{d_A^*}\ ,\\
		\frac{\Delta R_*}{R_*} &= \frac{\Delta d_A^*}{d_A^*} + \frac{\delta\omega_m}{2\omega_m}\ .
	\end{align}
	\end{subequations}
	Here we are using the short-hand notation $d_A^*\equiv d_A(z_*)$. The variation in
	these two parameters is only a small fraction of all the possible changes that any
	modified cosmology can produce in the CMB. If we want to compute all these changes
	and definitively establish the level of agreement of a given model with the CMB, we
	must resort to a Boltzmann code and perform the numerical computation. However, we 
	can argue that in order not to be directly excluded, any reasonable $\Lambda$CDM
	extension must keep $\theta_*$ and $R_*$ approximately fixed. Then, imposing
	$\Delta\theta_*, \Delta R_*\simeq 0$, we obtain the following system
	\begin{subequations}
	\begin{align}
		\left(I_{d_A}^*-I_{r_\text{s}}^*\right)\frac{\delta h}{h} 
			+ \left(J_{d_A}^*-J_{r_\text{s}}\right)\frac{\delta\omega_m}{\omega_m}
			&= \int\frac{\di x_z}{1+x_z}\left(R^*_{r_\text{s}}-R^*_{d_A}\right)\frac{\delta H}{H}\ ,\\
		I_{d_A}^*\frac{\delta h}{h} 
			+ \left(J_{d_A}^*+\frac{1}{2}\right)\frac{\delta\omega_m}{\omega_m}
			&= - \int^\infty_0\frac{\di x_z}{1+x_z}R^*_{d_A}\frac{\delta H}{H}\ .
	\end{align}
	\end{subequations}
	Solving the system we get the response functions for $h$ and $\omega_m$
	\begin{subequations}
	\begin{align}
		\frac{\delta h}{h} &= \int^\infty_0\frac{\di x_z}{1+x_z}\, \mathcal{R}_h(x_z)\frac{\delta H(x_z)}{H(x_z)}\ ,\\
		\frac{\delta \omega_m}{\omega_m} &= \int^\infty_0\frac{\di x_z}{1+x_z}\, \mathcal{R}_{\omega_m}(x_z)\frac{\delta H(x_z)}{H(x_z)}\ .
	\end{align}
	\end{subequations}
	where
	\begin{subequations}\label{eq:def_Rh_Rom}
	\begin{align}
		\mathcal{R}_h &= \frac{1}{D_*}\left\{\left(J_{d_A}^*+\frac{1}{2}\right)
			\left(R_{r_\text{s}}(x, z_*)-R_{d_A}(x,z_*)\right) 
			- (I^*_{r_\text{s}}-I^*_{d_A})R_{d_A}(x, z_*)\right\}\ ,\\
		\mathcal{R}_{\omega_m} &= \frac{1}{D_*}\Big\{I^*_{r_\text{s}}R_{d_A}(x,z_*)
			- I_{d_A}^*R_{r_\text{s}}(x, z_*)\Big\}\ ,\\
		D^* &= I^*_{d_A}\left(J^*_{r_\text{s}}-J^*_{d_A}\right)
			-\left(I^*_{r_\text{s}}-I^*_{d_A}\right)\left(J^*_{d_A}+\frac{1}{2}\right)\ .
	\end{align}
	\end{subequations}
	The response functions allow us to connect the changes produced in the expansion
	history by a generic model with changes of the observationally preferred parameters
	in the new cosmology. The observations considered in this case are the CMB priors, 
	which ensure that all the modified cosmologies considered are roughly compatible
	with the CMB. While the expressions derived so far are general, we will restrict
	ourselves to late-time modifications, so that $\delta H(z)=0$ for $z>300$.
	
	We can now use the previous results to compute the response function of any
	other cosmological quantity
	\begin{equation}
		\frac{\Delta\mathcal{G}(z)}{\mathcal{G}(z)} =
			\int^\infty_0\frac{\di x_z}{1+x_z}\, \mathcal{R}_\mathcal{G}(x_z, z)\frac{\delta H(x_z)}{H(x_z)}\ ,
	\end{equation}
	where the response function $\mathcal{R}_\mathcal{G}$ can be expressed as
	\begin{equation}
		\mathcal{R}_\mathcal{G}(x_z, z) \equiv I_\mathcal{G}(z)\mathcal{R}_h(x_z)
			+ J_\mathcal{G}(z)\mathcal{R}_{\omega_m}(x_z) + R_\mathcal{G}(x_z, z)\ .
	\end{equation}
	These results allow us to answer one of the main questions of this work. 
	The response function $\mathcal{R}_h$, depicted in Figure \ref{fig:responses},
	is strictly negative, so to increase the value of $h$ and thus solve the Hubble
	tension we need $\delta H(z)<0$ for some $z$. In the context of dark energy models,
	according to \eqref{eq:Omegaz_w}, this means that the equation of state must be
	phantom-like, i.e. $w(z)<-1$ for some $z$. To reach this conclusion we only used 
	the fact that $\mathcal{R}_h$ is strictly negative. Its shape will be important in
	the next section, where we will try to simultaneously solve the $H_0$ and the 
	$\sigma_8$ tensions.
	
	Finally, the response function of $\omega_m$ is very close to zero in the whole range 
	$0<z<300$. Even though we will present the analytical results with full
	generality, for late-time modifications it is completely justified to keep 
	$\omega_m$ fixed, i.e. $\mathcal{R}_{\omega_m}\to 0$. The variation of the Hubble
	parameter can then be obtained using only the first CMB prior in \eqref{eq:CMBpriors_variation}
	\begin{equation}
		\frac{\delta h}{h} \simeq -\frac{1}{I^*_{d_A}}\int^\infty_0
			\frac{\di x_z}{1+x_z}R^*_{d_A}\frac{\delta H}{H}\ .
	\end{equation}

\section{Growth factor and the $\sigma_8$ tension}\label{sec:sigma8}
	\subsection{Growth factor and $\sigma_8$ in $\Lambda$CDM}
		After decoupling, the time evolution of matter perturbations can be 
		encapsulated in the growth factor. The growth factor in $\Lambda$CDM obeys
		\begin{equation}\label{eq:evol_Da}
			\d{^2 D}{a^2} + \d{\log(a^3 H)}{a}\d{D}{a}
				-F(a)D= 0\ , \qquad F(a)\equiv \frac{3\Omega_m H^2_0}{2a^5H^2}\ .
		\end{equation}
		This equation remains valid even if the expansion history $H(a)$ is different
		from $\Lambda$CDM, as long as the equations describing the perturbations are not 
		modified. For a late-time $\Lambda$CDM universe, where matter and $\Lambda$ are 
		the	dominant components, the two independent solutions of
		\eqref{eq:evol_Da} can be expressed analitically
		\begin{subequations}\label{eq:analytic_Da}
		\begin{align}
			D_+(a) &= \frac{5\Omega_m}{2}\frac{H(a)}{H_0}I(a)\ ,\qquad 
				I(a)\equiv \int^a_0\di x_a\frac{H_0^3}{\big(x_aH(x_a)\big)^3}\ ,\\
			D_-(a) &\propto H(a)\ ,
		\end{align}
		\end{subequations}
		where $D_+$ and $D_-$ are the growing and decaying mode, respectively.
		It is also common to define the linear growth rate $f$, that in $\Lambda$CDM
		can be approximated as
		\begin{equation}
			f\equiv \d{\log{D_+}}{\log a} \simeq \left(\frac{\Omega_m H_0^2a^{-3}}{H^2}\right)^{0.55}.
		\end{equation}
		We start with the
		definition, e.g. see \cite{dodelson2020modern},
		\begin{equation}\label{eq:def_sigmaR_2}
			\sigma_R^2\equiv \av{\delta^2_{m,R}(\v{x})}\ ,
		\end{equation}
		where
		\begin{equation}
			\delta_{m,R}(\v{x})\equiv \int\di^3x'\,\delta_{m}(\v{x}')W_R\big(|\v{x}-\v{x}'|\big)\ ,
			\qquad W_R\big(r\big)=
				\left\{\begin{array}{ll}
					\displaystyle\frac{3}{4\pi R^3}\ , &\quad x<R\\[8pt]
					0\ ,                               &\quad x> R
				\end{array}\right.			
		\end{equation}
		It is common practice to evaluate this averaged clustering amplitude in spheres
		of radius $R=8\ h^{-1}\text{Mpc}$ and denote it as $\sigma_8$. It can be equivalently
		expressed in Fourier space and in terms of the matter power spectrum as
		\begin{equation}\label{eq:def_sigmaR_2}
			\sigma_R^2 = \int\frac{\di k}{k}\mathcal{P}_m(k)W^2(kR)\ ,\qquad
				W(x) \equiv \frac{3j_1(x)}{x}\ ,
		\end{equation}
		where $j_1$ is a spherical Bessel function. 		
		The approximate form of the linear matter power spectrum in terms of the matter 
		growth factor and the transfer function is \cite{dodelson2020modern}
		\begin{equation}
			\mathcal{P}_m(k)\equiv \frac{k^3}{2\pi^2}P_m(k) = \frac{4}{25}\frac{k^4}{\Omega_m^2H_0^4}
				T^2(k)D^2_+(a)\mathcal{P}_\mathcal{R}(k)\ ,
		\end{equation}
		where the primordial power spectrum of curvature perturbations is
		\begin{equation}
			\mathcal{P}_\mathcal{R}(k)= A_s\left(\frac{k}{k_p}\right)^{n_s-1},\qquad
				k_p=0.05\,\text{Mpc}^{-1}\ .
		\end{equation}
		For the transfer function we will adopt the Eisenstein-Hu fitting formula \cite{Eisenstein:1997ik} 
		that takes into account the baryonic supression at small
		scales and proves important for an accurate computation of $\sigma_8$. Following
		the notation of the original work \cite{Eisenstein:1997ik}
		\begin{alignat}{2}
			q &= \frac{k}{\text{Mpc}^{-1}}\frac{\Theta_{2.7}^2}{\Gamma_\text{eff}}\ ,	&\qquad\qquad 
				\Gamma_\text{eff} &= \omega_m\left(\alpha_\Gamma + \frac{1-\alpha_\Gamma}{1+(0.43ks)^4}\right)\ ,\nonumber\\
			T_0(q) &= \frac{L_0}{L_0+C_0q^2}\ , &\qquad\qquad
				\alpha_\Gamma &= 1-0.328\log(431\omega_m)\frac{\omega_b}{\omega_m}
					+ 0.38\log(22.3\omega_m)\left(\frac{\omega_b}{\omega_m}\right)^2\ ,\nonumber\\
			L_0(q) &= \log(2\ee + 1.8 q)\ , &\qquad\qquad
				s &= \frac{44.5 \log(9.83/\omega_m)}{\sqrt{1+10(\omega_b)^{3/4}}}\,\text{Mpc}\ ,\nonumber\\
			C_0(q) &= 14.2 + \frac{731}{1+62.5q}\ ,
		\end{alignat}
		where $\Theta_{2.7}$ is the temperature of the CMB in $2.7\ \text{K}$ units.
		So finally, the transfer function that we will use is $T_\text{EH}(k) = T_0\big(q(k)\big)$ 
		Also notice that we will always assume that $k$ in the integral is measured in 
		$\text{Mpc}^{-1}$ and not in $\text{Mpc}^{-1}\,h$ units.		
		After rewriting \eqref{eq:def_sigmaR_2}, we can write the $\sigma_8$ as
		\begin{equation}
			\sigma_8^2 = \frac{4}{25\omega_m}D_+^2(a)\mathcal{I}_k\ ,
		\end{equation}
		where
		\begin{equation}\label{eq:def_Ik}
			\mathcal{I}_k = \int^\infty_0\frac{\di k}{k}\left(\frac{k}{C_H}\right)^4
				T^2(k)W^2(kR)\mathcal{P}_\mathcal{R}(k)\ ,
		\end{equation}
		and again $R=8\ h^{-1}\text{Mpc}$.
		Closely related, $S_8$ is defined as
		\begin{equation}
			S_8 \equiv \sigma_8\sqrt{\frac{\Omega_m}{0.3}}\ .
		\end{equation}
		This quantity is closer to what is actually measured in weak-lensing surveys and
		is commonly used to reformulate the $\sigma_8$ tension as a $S_8$ tension. 
		Spectroscopic surveys on the other hand usually target the combination $f\sigma_8$,
		that can be precisely measured with redshift-space distortions. 
		
	\subsection{The $\sigma_8$ tension}
		The evolution of the variation of the growth factor is described by
		\begin{equation}\label{eq:evol_DeltaDa}
			\d{^2}{a^2}\Delta D + \d{\log(a^3H)}{a}\d{}{a}\Delta D -F(a)\Delta D = g(a) \ ,
		\end{equation}
		where 
		\begin{equation}
			g(a)\equiv -\d{}{a}\left(\frac{\Delta H}{H}\right)\d{D}{a}
				+ FD\left(\frac{\delta\omega_m}{\omega_m}-2\frac{\Delta H}{H}\right)\ .
		\end{equation}
		Using the Wronskian method, we can construct the particular solution to the 
		inhomogeneous equation \eqref{eq:evol_DeltaDa} and express the variations of
		the growth factor and the linear growth rate as
		\begin{align}
			\Delta D &= \frac{H(a)}{H_0} \int^a_0\di x_a \frac{x_a^3H^2(x_a)}{H_0^2}\Big(I(a)-I(x_a)\Big) g(x_a)\ ,\label{eq:DeltaD}\\
			\Delta f &= \d{}{\log a}\frac{\Delta D}{D}\ .
		\end{align}
		The full analytical expressions for the $(I, J, R)$ pieces of the variations
		can be found in Appendix \ref{sec:app_formulae}. We are now in position to 
		compute the variation in the $\sigma_8$ clustering amplitude. 		
		This variation can be written as the combination
		\begin{equation}
			\frac{\Delta\sigma_8}{\sigma_8} = \frac{\Delta D}{D} 
				- \frac{\delta\omega_m}{\omega_m} 
				+ \frac{1}{2}\frac{\Delta \mathcal{I}_k}{\mathcal{I}_k}\ .
		\end{equation}
		We still need to compute the variations on the integral $\mathcal{I}_k$
		\begin{align}
			\Delta \mathcal{I}_k &= 2\int^\infty_0\frac{\di k}{k}T^2(k)\left(\frac{k}{C_H}\right)^4W(kR)\Delta W(kR)\mathcal{P}_\mathcal{R}(k)\nonumber\\
				&\quad + 2\int^\infty_0\frac{\di k}{k} T(k)\Delta T(k)\left(\frac{k}{C_H}\right)^4W^2(kR)\mathcal{P}_\mathcal{R}(k)\ .
		\end{align}
		Similarly, the variation of $S_8$ is
		\begin{equation}
			\frac{\Delta S_8}{S_8} = \frac{\Delta\sigma_8}{\sigma_8} - \frac{\delta h}{h} + \frac{1}{2}\frac{\delta\omega_m}{\omega_m}\ .
		\end{equation}
		
		The response functions for $\sigma_8$, $f\sigma_8$ and $S_8$ are represented in
		Figure \ref{fig:responses}. Both $\mathcal{R}_{\sigma_8}$ and $\mathcal{R}_{f\sigma_8}$
		are strictly negative, which means that in order to reduce them we need $\delta H(z)>0$
		at some $z$. This can be compared with the result of the previous section, which
		showed that in order to increse $H_0$ we need $\delta H(z)<0$. The bottom line of
		this analysis is that both conditions must be fulfilled to solve the two cosmological
		tensions, otherwise we improve one at the cost of worsening the other. In particular,
		for a dark energy model \eqref{eq:Omegaz_w} a change of sign in $\delta H(z)$ implies
		that the equation of state $w(z)$ must cross the value $w=-1$. However, the results
		for $S_8$ are slightly different, since at very late-times the response function
		changes its sign. This feature could be very positive if,
		with the results of upcoming LSS surveys, we find ourselves
		in a situation where the clustering amplitude tension is clearly more severe in 
		$S_8$ or in $f\sigma_8$. The different behaviour of their response functions might
		be then a clear explanation and could give us hints about the shape of 
		$\delta H(z)$.
		
		\begin{figure}[ht]
			\centering
			\begin{subfigure}[t]{0.48\textwidth}
				\includegraphics[scale=0.5]{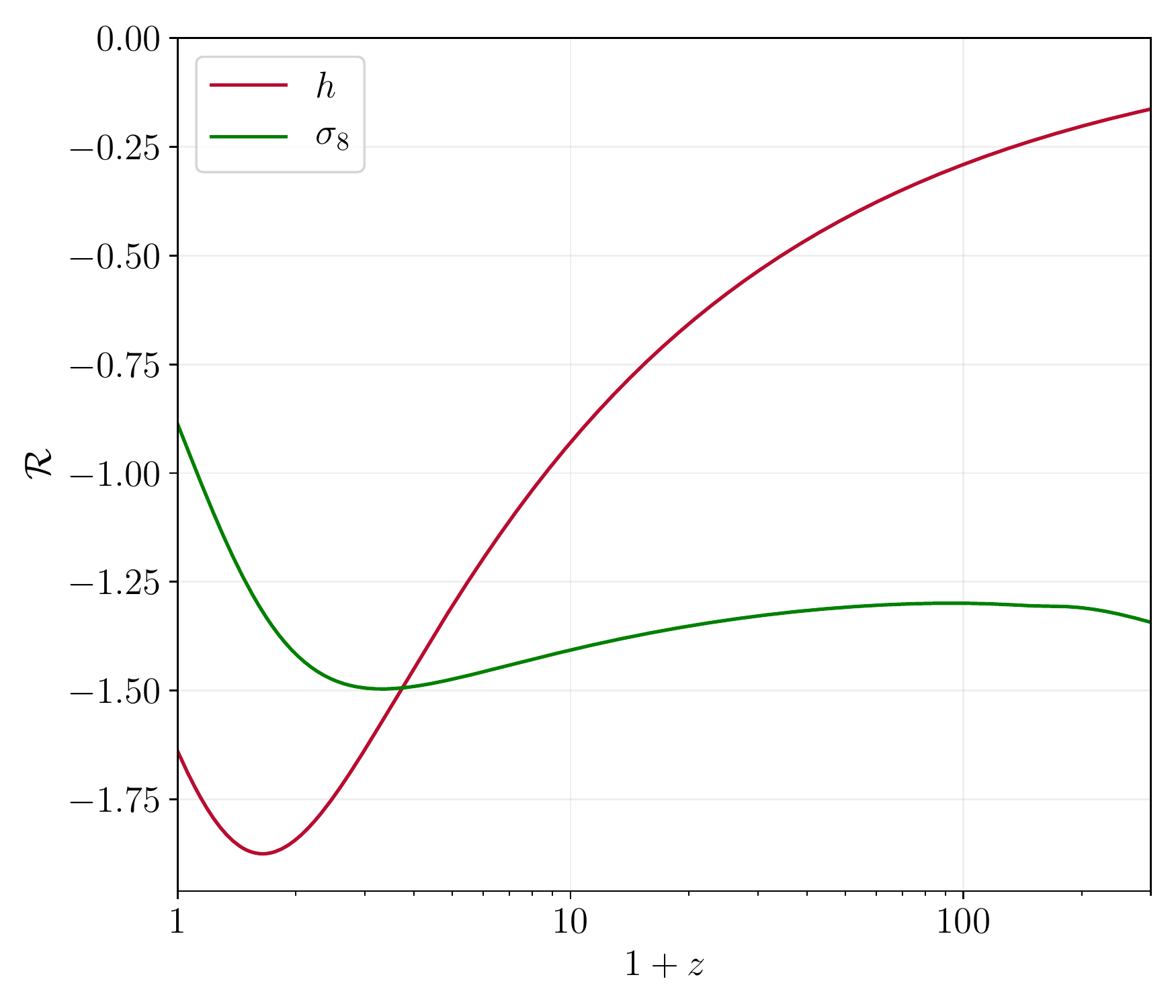}
			\end{subfigure}
			\begin{subfigure}[t]{0.48\textwidth}
				\includegraphics[scale=0.5]{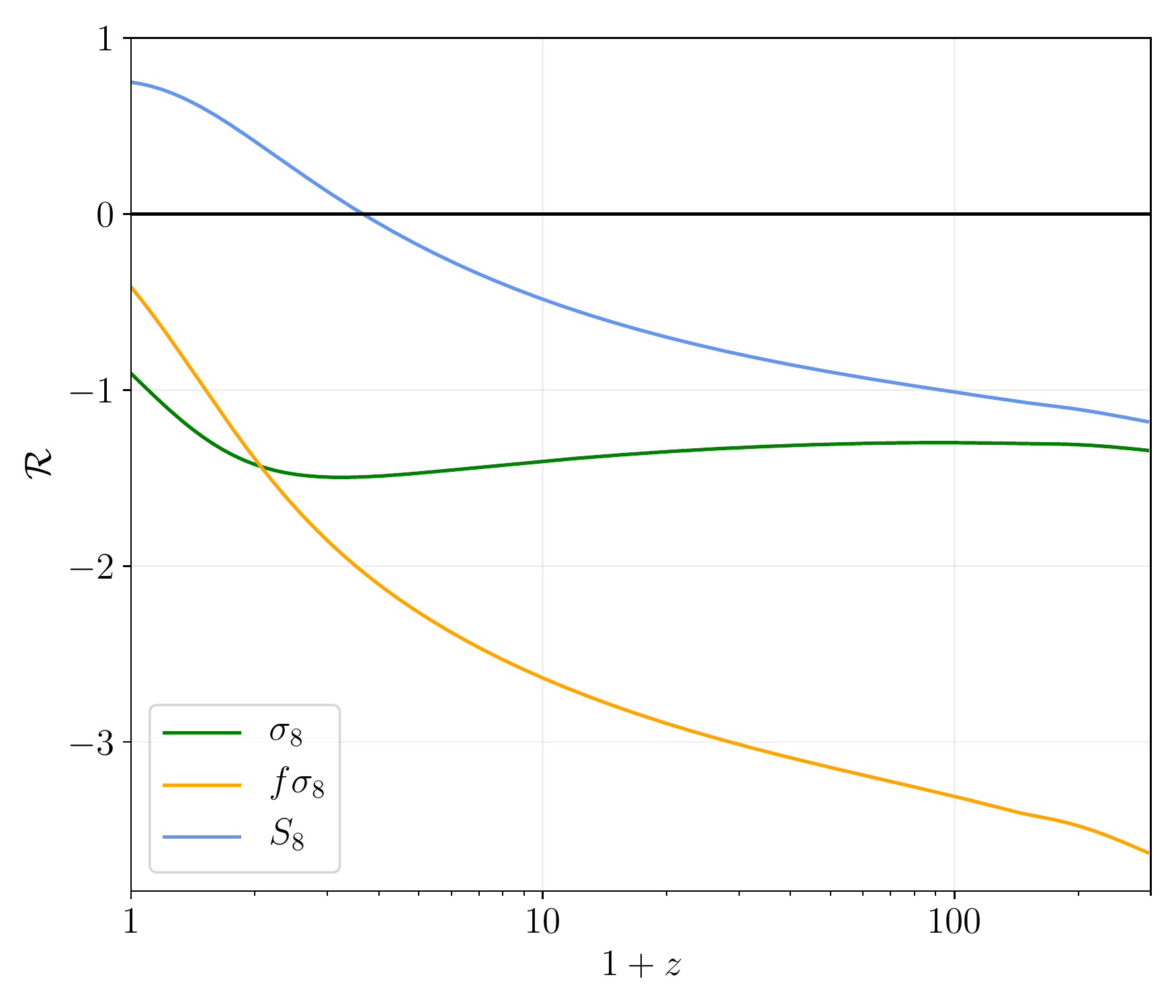}
			\end{subfigure}
			\caption{(Left) Response functions for $h$ and $\sigma_8$. Notice that
				both have the same sign so, unless $\delta H$ changes sign, both
				variations follow the same trend, i.e. if we increase $h$ we also
				increase $\sigma_8$. (Right) Response functions for other 
				clustering-related quantities.}
			\label{fig:responses}
    	\end{figure}

	\subsection{Deformations beyond the background: $G_\text{eff}$}\label{sec:G_eff}
		We define $G_\text{eff}$ as a modification in the sub-Hubble regime that
		leads to the modified evolution for the growth factor 
		\begin{equation}
			\d{^2D}{a^2} + \d{\log (a^3H)}{a}\d{D}{a} - \frac{G_\text{eff}}{G}F(a)D = 0\ .
		\end{equation}
		Many realistic scenarios actually produce this kind of modification, e.g. see
		\cite{Heisenberg:2020xak}.
		Following the same steps as in previous sections, if we assume that the effective
		gravitational coupling is close to the $\Lambda$CDM case, $G_\text{eff}=G+\delta G(z)$,
		we get 
		\begin{equation}
			\d{^2}{a^2}\Delta D + \d{\log (a^3H)}{a}\d{}{a}\Delta D 
				- F(a)\Delta D = F(a)D(a)\frac{\delta G(a)}{G}\ ,
		\end{equation}
		where in this case $\Delta D$ stands for a variation keeping fixed all the 
		cosmological parameters and $H(z)$.	The particular solution is 
		\begin{align}
			\frac{\Delta D}{D} &= \frac{H(a)}{D(a)H_0}\int^a_0\di x_a \frac{x_a^3H^2(x_a)}{H_0^2}\Big(I(a)-I(x_a)\Big)F(x_a)D(x_a)\frac{\delta G(x_a)}{G}\nonumber\\
				&= \int^\infty_0 \frac{\di x_z}{1+x_z} \mathcal{G}(x_z, z)\frac{\delta G(x_z)}{G}\ ,
		\end{align}
		where the response function for $\delta G$ is
		\begin{equation}\label{eq:G_eff_response}
			\mathcal{G}(x_z, z) = \frac{H^3(z) D(x_z)}{H_0^3 D(z)}\Big(I(z)-I(x_z)\Big)F(x_z)\frac{\theta(x_z-z)}{(1+x_z)^4}\ .
		\end{equation}
		Since we are working to first order, the complete variation, modifying the 
		background as well, is just a linear combination with the results of the 
		previous section, i.e.
		\begin{equation}
			\frac{\Delta D}{D}\bigg\lvert_\text{full} = \int^{\infty}_0\frac{\di x_z}{1+x_z}\left(
				\mathcal{R}_D(x_z, z)\frac{\delta H(x_z)}{H(x_z)}
				+\mathcal{G}(x_z, z)\frac{\delta G(x_z)}{G}\right)\ .
		\end{equation}
		
		Including two free functions $\delta H(z)$ and $\delta G(z)$ make the results
		more general but unfortunately prevent us from making strong statements about
		the behaviour of any of them. In order to proceed further, we restrict ourselves
		to the case in which $\delta H(z)$ does not change sign. We know that this scenario
		is realized in many physically relevant models, for instance when we have a dark
		energy fluid that does not cross the phantom divide. 
		
		We already discussed that in this case in order to solve the $H_0$ tension we
		require $\delta H<0$. If we do not modify the evolution of the perturbations, 
		i.e. $G_\text{eff}=G$, this leads to an increase in $\sigma_8$, since
		$\mathcal{R}_{\sigma_8}<0$. However, including $\delta G(z)$ we have enough
		freedom to increase $H_0$ while reducing $\sigma_8$. Intuitively, it seems
		evident that we can achieve this goal just reducing the effective strength
		of gravity enough, i.e. $\delta G(z)<0$. Defining $\alpha(x_z)\equiv
		-\mathcal{R}_{\sigma_8}(x_z, 0)/\mathcal{G}(x_z, 0)$ and using the previous results 
		we can derive the stronger condition
		\begin{equation}\label{eq:G_eff_bound}
			\frac{\delta G(x_z)}{G} < \alpha(z)\frac{\delta H(x_z)}{H(x_z)}<0\qquad
				\text{for some }x_z>0\ ,
		\end{equation}
		that a model must satisfy if we want to reduce the value of $\sigma_8$, while 
		increasing $H_0$.

\section{Supernova absolute magnitude}\label{sec:supernovae}
	Most discussions on the Hubble tension are formulated in terms of the $H_0$, or $h$, 
	parameter. However, the parameter that is closer to what is actually measured
	by collaborations like SH0ES \cite{Riess:2019cxk}, is the absolute magnitude $M$
	used to calibrate the observed apparent magnitudes of SNe. This is the actual source
	of the Hubble tension, as has been stressed by \cite{Camarena:2021jlr}, which
	also presented models where $H_0$ is raised without affecting $M$. In this section
	we will compute the response function for $M$, paying special attention to its 
	differences with respect to the response function for $h$.
	The apparent magnitude and distance modulus are defined as \cite{Pan-STARRS1:2017jku}
	\begin{align}
		m &\equiv 5\log_{10}\left(\frac{d_L}{\text{Mpc}}\right) + 25 + M\ ,\\
		\mu &\equiv m - M\ ,
	\end{align}
	where $M$ is the absolute magnitude, that must be calibrated to infer the distance
	from the observed apparent magnitude. The $\chi^2$ can then be constructed as
	\begin{equation}
		\chi^2_\text{\tiny{SNe}} = (m_\text{obs}^i-m_\text{th}^i)(C^{-1})_{ij}(m_\text{obs}^j-m_\text{th}^j)\ ,
	\end{equation}
	and it can be analitically minimized for $M$
	\begin{equation}
		\pd{\chi^2_\text{\tiny{SNe}}}{M} = 0\qquad\to\qquad 
			M_\text{best\ fit} = \frac{\sum_{ij}(C^{-1})_{ij}(m_\text{obs}^j-\mu_\text{th}^j)}{\sum_{ij}(C^{-1})_{ij}}\ .
	\end{equation}
	For given values of $H_0$ and $\Omega_m$, this is the absolute magnitude
	that provides the best fit to SNe data. Its variation is
	\begin{equation}
		\Delta M = -5\frac{\delta h}{h} - \frac{5}{\sum_{ij}(C^{-1})_{ij}}\sum_{ij}\frac{(C^{-1})_{ij}\Delta d_L(z_j)}{d_L(z_j)}\ ,
	\end{equation}
	so we have
	\begin{equation}\label{eq:R_M}
		\mathcal{R}_M(x_z) = -5\mathcal{R}_h(x_z) - \frac{5}{\sum_{ij}(C^{-1})_{ij}}\sum_{ij}(C^{-1})_{ij}\mathcal{R}_{d_L}(x_z, z_j)\ .
	\end{equation}	
	In this work, we will use the Pantheon sample \cite{Pan-STARRS1:2017jku} for the computation 
	of \eqref{eq:R_M}. In Figure \ref{fig:responses}, we can see that, in contrast with $h$, 
	the response function for $M$ changes sign at low redshift. This means that models that 
	rely on modifications at very low redshift may increase the Hubble constant without 
	actually decreasing $M$, a result in line with the conclusions of \cite{Camarena:2021jlr}. 
	
	\begin{figure}[ht]
		\centering
		\begin{subfigure}[t]{0.48\textwidth}
			\includegraphics[scale=0.5]{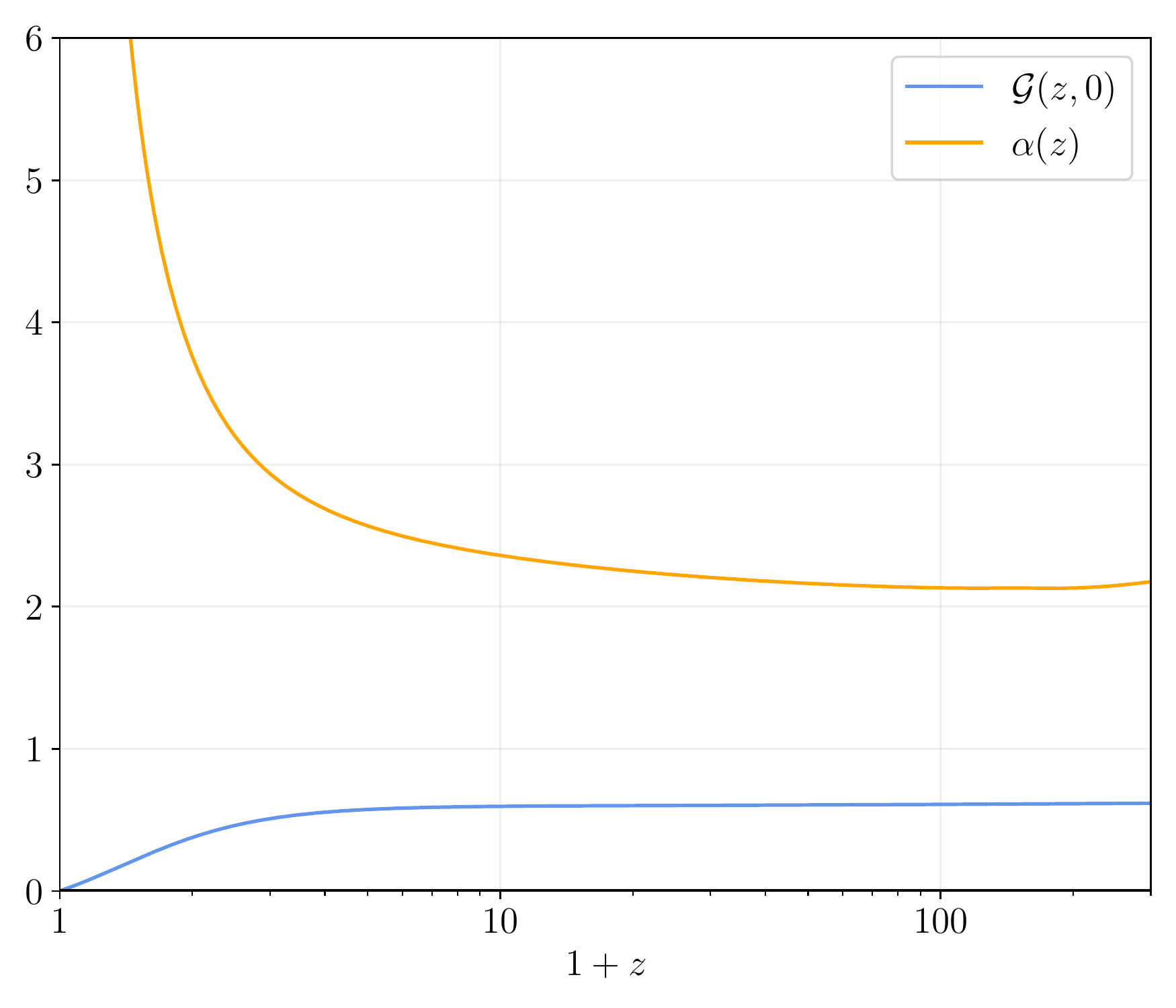}
		\end{subfigure}
		\begin{subfigure}[t]{0.48\textwidth}
			\includegraphics[scale=0.5]{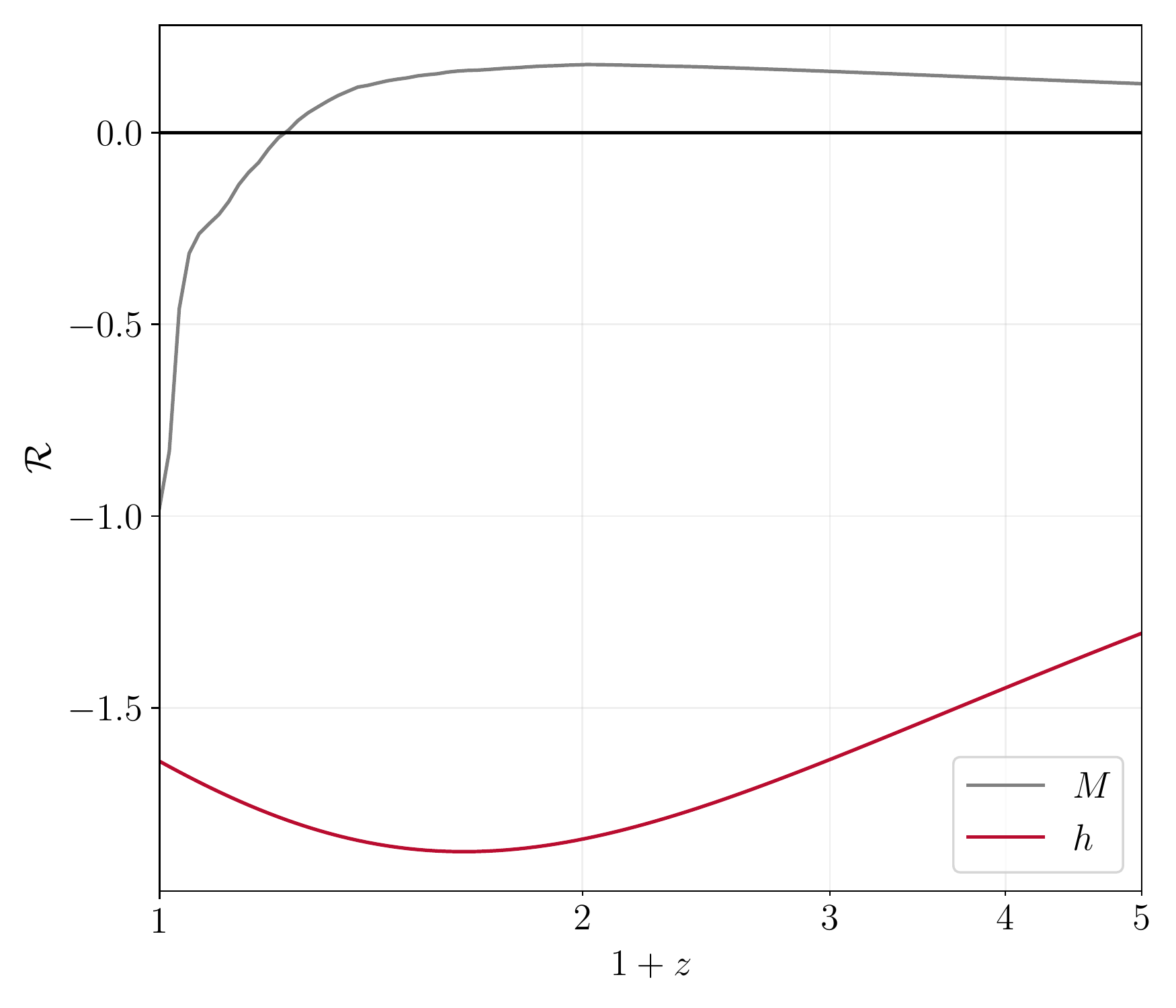}
		\end{subfigure}
		\caption{(Left) Response of $\sigma_8$ to small changes in the effective 
			gravitational constant, $\mathcal{G}$ in \eqref{eq:G_eff_response}. The second
			curve $\alpha(z)$ depicts the bound defined in \eqref{eq:G_eff_bound}.
			(Right) Response function of $h$ and the supernova absolute magnitude $M$. 
			Notice that the latter becomes negative at very low redshift, so a very late-time 
			modification that tries to increase the value of $h$ would also
			increase $M$ and thus fail at addressing the $H_0$ tension
			between the CMB and supernovae-based direct measurements.}
		\label{fig:responses}			
   	\end{figure}

\section{Summary and conclusions}\label{sec:summary}

In this paper we have addressed the question of why typical late-time dark energy models 
only solve the $H_0$ tension at the cost of predicting a large clustering amplitude 
$\sigma_8$ and whether it is therefore actually possible to relieve both tensions 
simultaneously by perturbatively modifying the expansion history, and maybe the
gravitational constant, at late times. Using a model independent approach we derived a set 
of necessary conditions on the functional form of $\delta H(z)$ which have to be satisfied 
in order to tackle both the $H_0$ and $\sigma_8$ tensions. For the particularly interesting 
case in which the deformation is due to dark energy with equation of state $w(z)$ our 
results can be summarized schematically as follows
	\begin{enumerate}[i)]
        \item Solving the $H_0$ tension $\quad\Rightarrow\quad$ $\delta H(z)<0$ 
        	for some $z$ $\quad\Rightarrow\quad$ $w(z)<-1$ for some $z$.
		\item If the perturbations are not modified ($G_\text{eff}=G$) then:\\[3pt]
			Solving the $H_0$ and $\sigma_8$ tensions $\quad\Rightarrow\quad$
				$\delta H(z)$ changes sign $\quad\Rightarrow\quad$
                	$w(z)$ crosses the phantom divide.
        \item If $G_\text{eff} = G +\delta G(z)$ and $\delta H(z)$ does not change sign then:\\[3pt]
        	Solving the $H_0$ and $\sigma_8$ tensions $\quad\Rightarrow\quad$
        	$\displaystyle\frac{\delta G(z)}{G}<\alpha(z)\frac{\delta H(z)}{H(z)}<0\quad$  for some $z$.\\[3pt]
        	where $\delta H(z)<0$ and $\alpha(z)>0$.
        \item Solutions that rely on significant modifications at low redshift ($z<1$) can
        	increase $H_0$ without decreasing the supernova absolute magnitude $M$, thus
        	failing to adress the Hubble tension.
	\end{enumerate}
	
Note that while we chose here to present the implications of our results for the specific 
case of a dark energy model, the conditions on the form of $\delta H(z)$ are much more 
general and can be applied to any theory.
	
Providing a full catalog of models ruled out by these necessary criteria will be left for 
future work, as well as the further study of theories meeting them. It would for example be 
interesting to include low redshift constraints from baryon acoustic oscillation (BAO) data 
and address concerns along the lines of \cite{Benevento:2020fev}.

Another interesting avenue would be to extend this results to early dark energy models.
In this case, the CMB would be modified in a different way and different observational 
anchors should be used. The variation of $\omega_m$, negligible for this work, would
have to be taken into account and would potentially play an important role.

Finally, while the computations in the present paper, in particular the solution \eqref{eq:DeltaD}, 
explicitly assume a $\Lambda$CDM background with only matter and curvature contributions on 
top of the cosmological constant, it is possible to generalize our analysis to arbitrary 
backgrounds. This will allow the application of the method to deviations from general DE 
models or theories beyond Einstein gravity and will be useful especially in the context of 
effective field theory of dark energy to consider observables at the perturbation level 
beyond $G_\text{eff}$ presented here.

\begin{acknowledgments}
	LH is supported by funding from the European Research Council (ERC) under the European Unions Horizon 2020 research and innovation programme grant agreement No 801781 and by the Swiss National Science Foundation grant 179740. 
	HVR is supported by the Spanish Ministry of Universities through a Margarita Salas Fellowship, with funding from the European Union under the NextGenerationEU programme.
\end{acknowledgments}

\appendix
\section{Full analytical results}\label{sec:app_formulae}
	This appendix contains the full analytical expressions used in this work. 
	Unless otherwise stated, every function inside the integrals depends on the integration 
	variable $x_z$.
	Comoving, luminosity and angular diameter distance:
	\begin{equation}
		\left\{\begin{array}{l}
		\displaystyle
			I_{\chi}(z) = I_{d_L}(z) = I_{d_A}(z) = -\frac{1}{\chi(z)}\int^z_0\di x_z\frac{H_0^2}{H^3}\\[8pt]
			\displaystyle
			J_{\chi}(z) = J_{d_L}(z) = J_{d_A}(z) = -\frac{1}{\chi(z)}\int^z_0\di x_z\frac{H_0^2}{H^3}m(x_z)\\[8pt]
			\displaystyle
			R_{\chi}(x_z,z) = R_{d_A}(x_z,z) = R_{d_A}(x_z,z) = -(1+x_z)\frac{\theta(z-x_z)}{\chi(z)H(x_z)}
		\end{array}\right.
	\end{equation}
	Comoving sound horizon:
	\begin{equation}
		\left\{\begin{array}{l}
		\displaystyle
			I_{r_\text{s}}(z) = -\frac{1}{r_\text{s}(z)}\int^\infty_z\di x_z\frac{H_0^2}{H^3}c_\text{s}(x_z)\\[8pt]
			\displaystyle
			J_{r_\text{s}}(z) = -\frac{1}{r_\text{s}(z)}\int^\infty_z\di x_z\frac{H_0^2}{H^3}m(x_z)c_\text{s}(x_z)\\[8pt]
			\displaystyle
			R_{r_\text{s}}(x_z,z) = -\frac{(1+x_z)c_\text{s}(x_z)}{r_\text{s}(z)}\frac{\theta(x_z-z)}{H(x_z)}
		\end{array}\right.
	\end{equation}
	Integral defined in \eqref{eq:def_Ik}:
	\begin{equation}
		\left\{\begin{array}{l}
			\displaystyle
			I_{\mathcal{I}_k} = -\frac{2}{\mathcal{I}_k}\int^\infty_0\frac{\di k}{k}T^2(k)\mathcal{P}_\mathcal{R}(k)\left(\frac{k}{C_H}\right)^4kR\,W(kR)W'(kR)\\[8pt]
			\displaystyle
			J_{\mathcal{I}_k} = \frac{2\omega_m}{\mathcal{I}_k}\int^\infty_0\frac{\di k}{k}T(k)\pd{T(k)}{\omega_m}\mathcal{P}_\mathcal{R}(k)\left(\frac{k}{C_H}\right)^4W^2(kR)\\[8pt]
			\displaystyle
			R_{\mathcal{I}_k} = 0
		\end{array}\right.
	\end{equation}
	Growth factor:
	\begin{equation}
		\left\{\begin{array}{l}
			\displaystyle
			I_{D}(z) = -\frac{H(z)}{H_0D(z)}\int^\infty_z\frac{\di x_z}{1+x_z}\frac{H_0^2D}{H^2}
				\Bigg\{\frac{H_0}{H}f + \frac{N(z, x_z)F}{1+x_z}
				\left(1+\frac{(1+x_z)^3}{F}\d{\log H}{x_z}f\right)\Bigg\}\\[8pt]
			\displaystyle
			J_{D}(z) = \frac{1}{5} + \frac{H(z)}{H_0D(z)}\int^\infty_z\frac{\di x_z}{(1+x_z)^2}N(z,x_z)FD \\
				\displaystyle \qquad\qquad
					- \frac{H(z)}{H_0D(z)}\int^\infty_z\frac{\di x_z}{1+x_z}mD
					\Bigg\{\frac{H_0}{H}f + \frac{N(z,x_z)F}{1+x_z}\left(1+\frac{(1+x_z)^3}{F}\d{\log H}{x_z}f\right)\Bigg\}\\[8pt]
			\displaystyle
			R_{D}(x_z,z) = -\frac{H(z)D(x_z)}{H_0D(z)}
				\Bigg\{\frac{H_0}{H(x_z)}f(x_z) + \frac{N(z, x_z)F(x_z)}{1+x_z}
				\left(1+\frac{(1+x_z)^3}{F(x_z)}\d{\log H}{x_z}f(x_z)\right)\Bigg\}
				\theta(x_z-z)
		\end{array}\right.
	\end{equation}
	Linear growth rate:
	\begin{equation}
		\left\{\begin{array}{l}
			\displaystyle
			I_{f}(z) = -I_D(z)\left(1+\frac{1+z}{f}\d{\log H}{z}\right) - \frac{H_0^2}{H^2(z)}\\[8pt] \displaystyle\qquad\qquad
				-\frac{(1+z)^2H_0^2}{f(z)H^2(z)D(z)}\int^\infty_z\frac{\di x_z}{(1+x_z)^5} FD\left(1+\frac{(1+x_z)^3}{F}\d{\log H}{x_z}f\right)\\[8pt] \displaystyle
			J_{f}(z) = -\left(J_D(z)-\frac{1}{5}\right)\left(1+\frac{1+z}{f}\d{\log H}{z}\right) - m(z)\\[8pt] \displaystyle\qquad\qquad
				+\frac{(1+z)^2}{f(z)H^2(z)D(z)}\int^\infty_z\frac{\di x_z}{(1+x_z)^5} H^2FD
					\Big\{1-m\left(1+\frac{(1+x_z)^3}{F}\d{\log H}{x_z}f\right)\Big\}\\[8pt] \displaystyle
			R_{f}(x_z,z) = -R_D(x_z, z)\left(1+\frac{1+z}{f}\d{\log H}{z}\right) - \delta(x_z-z)\\[8pt] \displaystyle \qquad\qquad\qquad
				-\frac{1}{(1+z)f(z)}\frac{(1+z)^3H^2(x_z)D(x_z)}{(1+x_z)^3H^2(z)D(z)}\frac{F(x_z)}{1+x_z}\left(1+\frac{(1+x_z)^3}{F(x_z)}\d{\log H}{x_z}f(x_z)\right)\theta(x_z-z)
		\end{array}\right.
	\end{equation}
	Supernova absolute magnitude:
	\begin{equation}
		\left\{\begin{array}{l}
			\displaystyle
			I_{M} = -\frac{5}{M}-\frac{5}{M\sum_{ij}(C^{-1})_{ij}}\sum_{ij}(C^{-1})_{ij}I_\chi(z_j)\\[8pt]
			\displaystyle
			J_{M} = -\frac{5}{M\sum_{ij}(C^{-1})_{ij}}\sum_{ij}(C^{-1})_{ij}J_\chi(z_j)\\[8pt]
			\displaystyle
			R_{M}(x_z) = -\frac{5}{M\sum_{ij}(C^{-1})_{ij}}\sum_{ij}(C^{-1})_{ij}R_\chi(x_z, z_j)
		\end{array}\right.
	\end{equation}

\section{Numerical benchmarks}\label{sec:app_benchmarks}
	The analytical expressions of the previous section have been tested for two particular
	dark energy models, $w=\text{const.}$ and the CPL \cite{Chevallier:2000qy, Linder:2002et} 
	parameterization $w=w_0 + w_a(1-a)$. In the latter, we choose fix parameter $w_0=-1.05$
	to obtain a deformation $\delta H(z)$ that changes sign at late times. We compare
	the analytical results with the numerical ones obtained using \texttt{class} 
	\cite{Blas:2011rf}, keeping	fixed the acoustic scale $\theta_*$ and $\omega_m$. 
	The analytic results show a very satisfactory performance as can be seen in Tables
	\ref{tab:w0} and \ref{tab:w0wa}.
	\begin{figure}[ht]
		\centering
		\begin{subfigure}[t]{0.48\textwidth}
			\includegraphics[scale=0.55]{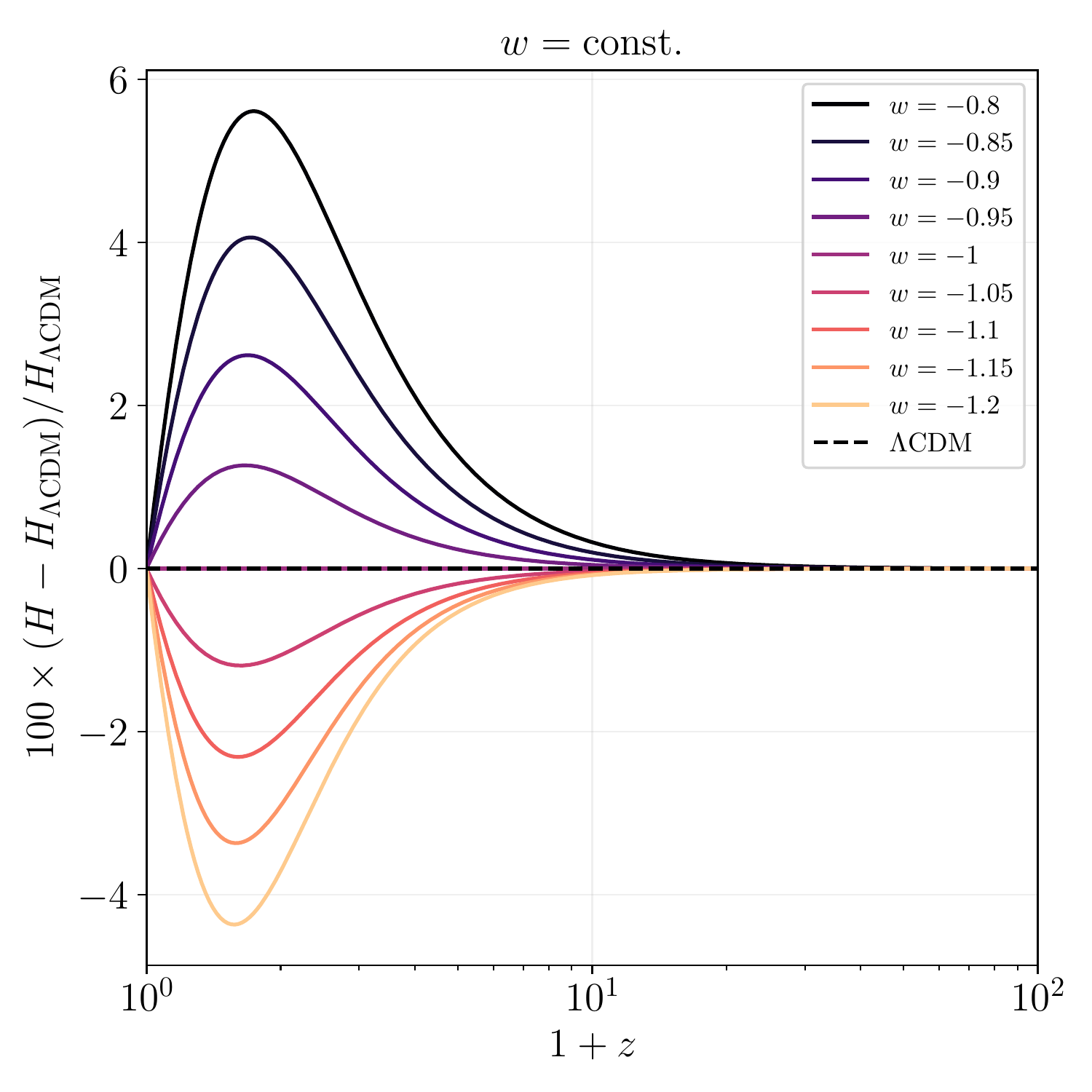}
		\end{subfigure}
		\begin{subfigure}[t]{0.48\textwidth}
			\includegraphics[scale=0.55]{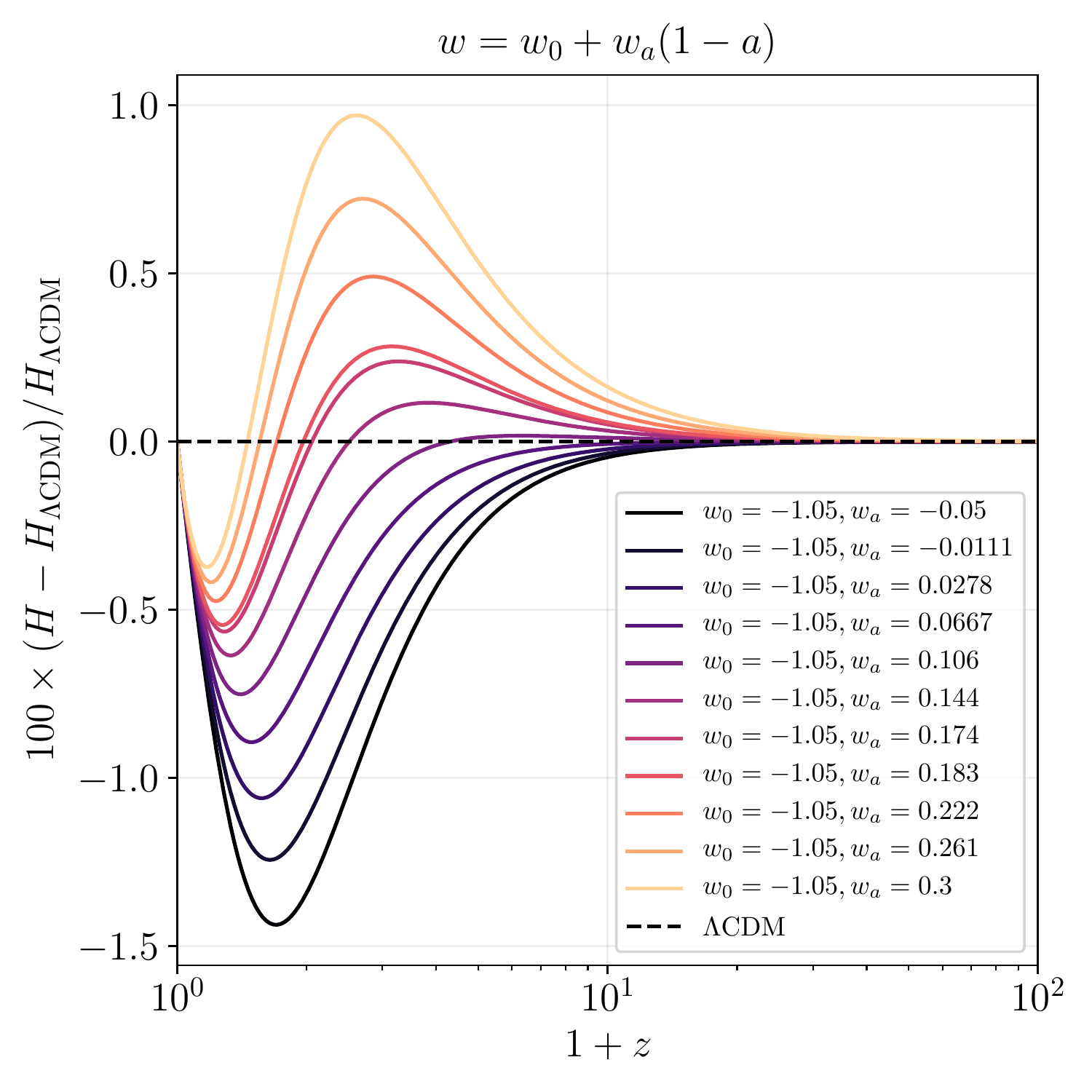}
		\end{subfigure}
		\caption{$\delta H(z)$ for two common dark energy parameterizations.}
		\label{fig:dH_w}			
    \end{figure}
    
    \begin{table}[ht]
		\begin{tabular}{r|r@{\hskip 12pt}rcr@{\hskip 12pt}r}
                                     & \multicolumn{2}{c}{$100\times\delta h/h$} 
                                     & 
                                     & \multicolumn{2}{c}{$100\times \Delta \sigma_8/\sigma_8$}\\[4pt]
                                $w$  & \texttt{class}
                                     & Analytical
                                     & \phantom{aaa}
                                     & \texttt{class}
                                     & Analytical\\
	    	\hline\hline
	    	$-0.80  $&$ -8.57   $&$   -10.97   $&&$   -8.98    $&$     -7.13$\\
			$-0.85 $&$ -6.47   $&$   -7.76    $&&$   -6.29    $&$     -5.32$\\
			$-0.90  $&$ -4.35   $&$   -4.89    $&&$   -3.93    $&$     -3.53$\\
			$-0.95 $&$ -2.19   $&$   -2.32    $&&$   -1.85    $&$     -1.76$\\
			$-1.05 $&$  2.22   $&$    2.10    $&&$    1.65    $&$      1.75$\\
			$-1.10  $&$  4.47   $&$    4.01    $&&$    3.12    $&$      3.48$\\
			$-1.15 $&$  6.75   $&$    5.74    $&&$    4.45    $&$      5.21$\\
			$-1.20  $&$  9.07   $&$    7.33    $&&$    5.66    $&$      6.93$
	    \end{tabular}
	    \caption{Comparison of our analytical results with the full computation
	    	in \texttt{class}, keeping fixed $\theta_*$ and $\omega_m$, for a
	    	dark energy model with a constant equation of state $w$.}
	    \label{tab:w0}
	\end{table}
				
	\begin{table}[ht]            
        \begin{tabular}{r|r@{\hskip 12pt}rcr@{\hskip 12pt}r}
            				         & \multicolumn{2}{c}{$100\times\delta h/h$} 
                                     & 
                                     & \multicolumn{2}{c}{$100\times \Delta \sigma_8/\sigma_8$}\\[4pt]
                              $w_a$  & \texttt{class}
                                     & Analytical
                                     & \phantom{aaa}
                                     & \texttt{class}
                                     & Analytical\\
	    	\hline\hline
	    	$-0.05   $&$   2.81   $&$    2.61    $&&$    2.24    $&$     2.08 $\\
			$-0.01   $&$   2.35   $&$    2.22    $&&$    1.86    $&$     1.75 $\\
			$ 0.03   $&$   1.89   $&$    1.80    $&&$    1.47    $&$     1.39 $\\
			$ 0.07   $&$   1.42   $&$    1.37    $&&$    1.07    $&$     1.03 $\\
			$ 0.11   $&$   0.946  $&$    0.93    $&&$    0.66    $&$     0.64 $\\
			$ 0.14   $&$   0.465  $&$    0.46    $&&$    0.24    $&$     0.23 $\\
			$ 0.174  $&$   0.095  $&$    0.093   $&&$   -0.089   $&$    -0.092$\\
			$ 0.18   $&$  -0.022  $&$   -0.025   $&&$   -0.19    $&$    -0.20 $\\
			$ 0.22   $&$  -0.52   $&$   -0.53    $&&$   -0.64    $&$    -0.65 $\\
			$ 0.26   $&$  -1.02   $&$   -1.07    $&&$   -1.09    $&$    -1.13 $\\
			$ 0.3    $&$  -1.54   $&$   -1.62    $&&$   -1.56    $&$    -1.64 $
	    \end{tabular}
	    \caption{Comparison of our analytical results with the full computation
	    	in \texttt{class}, keeping fixed $\theta_*$ and $\omega_m$, for a
	    	dark energy model with an equation of state $w(a)=-1.05+w_a(1-a)$. Notice
	    	that in this case $\delta H(z)$ changes sign, as shown in the right pannel in
	    	Figure \ref{fig:responses}, and for the case $w_a=0.174$ the variations
	    	have opposite signs. Even though they are too small to relieve the tensions, 
	    	this example shows that when $\delta H(z)$ changes sign it is possible to 
	    	increase $h$ while reducing $\sigma_8$.}
	    \label{tab:w0wa}
	\end{table}
	
\newpage
\bibliography{Biblio.bib}

\begin{thebibliography}{54}
\expandafter\ifx\csname natexlab\endcsname\relax\def\natexlab#1{#1}\fi
\expandafter\ifx\csname bibnamefont\endcsname\relax
  \def\bibnamefont#1{#1}\fi
\expandafter\ifx\csname bibfnamefont\endcsname\relax
  \def\bibfnamefont#1{#1}\fi
\expandafter\ifx\csname citenamefont\endcsname\relax
  \def\citenamefont#1{#1}\fi
\expandafter\ifx\csname url\endcsname\relax
  \def\url#1{\texttt{#1}}\fi
\expandafter\ifx\csname urlprefix\endcsname\relax\def\urlprefix{URL }\fi
\providecommand{\bibinfo}[2]{#2}
\providecommand{\eprint}[2][]{\url{#2}}

\bibitem[{\citenamefont{Aghanim et~al.}(2020)}]{Planck:2018vyg}
\bibinfo{author}{\bibfnamefont{N.}~\bibnamefont{Aghanim}} \bibnamefont{et~al.}
  (\bibinfo{collaboration}{Planck}), \bibinfo{journal}{Astron. Astrophys.}
  \textbf{\bibinfo{volume}{641}}, \bibinfo{pages}{A6} (\bibinfo{year}{2020}),
  \bibinfo{note}{[Erratum: Astron.Astrophys. 652, C4 (2021)]},
  \eprint{1807.06209}.

\bibitem[{\citenamefont{Riess et~al.}(2019)\citenamefont{Riess, Casertano,
  Yuan, Macri, and Scolnic}}]{Riess:2019cxk}
\bibinfo{author}{\bibfnamefont{A.~G.} \bibnamefont{Riess}},
  \bibinfo{author}{\bibfnamefont{S.}~\bibnamefont{Casertano}},
  \bibinfo{author}{\bibfnamefont{W.}~\bibnamefont{Yuan}},
  \bibinfo{author}{\bibfnamefont{L.~M.} \bibnamefont{Macri}}, \bibnamefont{and}
  \bibinfo{author}{\bibfnamefont{D.}~\bibnamefont{Scolnic}},
  \bibinfo{journal}{Astrophys. J.} \textbf{\bibinfo{volume}{876}},
  \bibinfo{pages}{85} (\bibinfo{year}{2019}), \eprint{1903.07603}.

\bibitem[{\citenamefont{Pesce et~al.}(2020)}]{Pesce:2020xfe}
\bibinfo{author}{\bibfnamefont{D.~W.} \bibnamefont{Pesce}}
  \bibnamefont{et~al.}, \bibinfo{journal}{Astrophys. J. Lett.}
  \textbf{\bibinfo{volume}{891}}, \bibinfo{pages}{L1} (\bibinfo{year}{2020}),
  \eprint{2001.09213}.

\bibitem[{\citenamefont{Wong et~al.}(2020)}]{Wong:2019kwg}
\bibinfo{author}{\bibfnamefont{K.~C.} \bibnamefont{Wong}} \bibnamefont{et~al.},
  \bibinfo{journal}{Mon. Not. Roy. Astron. Soc.}
  \textbf{\bibinfo{volume}{498}}, \bibinfo{pages}{1420} (\bibinfo{year}{2020}),
  \eprint{1907.04869}.

\bibitem[{\citenamefont{Riess et~al.}(2021)\citenamefont{Riess, Casertano,
  Yuan, Bowers, Macri, Zinn, and Scolnic}}]{Riess:2020fzl}
\bibinfo{author}{\bibfnamefont{A.~G.} \bibnamefont{Riess}},
  \bibinfo{author}{\bibfnamefont{S.}~\bibnamefont{Casertano}},
  \bibinfo{author}{\bibfnamefont{W.}~\bibnamefont{Yuan}},
  \bibinfo{author}{\bibfnamefont{J.~B.} \bibnamefont{Bowers}},
  \bibinfo{author}{\bibfnamefont{L.}~\bibnamefont{Macri}},
  \bibinfo{author}{\bibfnamefont{J.~C.} \bibnamefont{Zinn}}, \bibnamefont{and}
  \bibinfo{author}{\bibfnamefont{D.}~\bibnamefont{Scolnic}},
  \bibinfo{journal}{Astrophys. J. Lett.} \textbf{\bibinfo{volume}{908}},
  \bibinfo{pages}{L6} (\bibinfo{year}{2021}), \eprint{2012.08534}.

\bibitem[{\citenamefont{Abbott et~al.}(2018)}]{DES:2017myr}
\bibinfo{author}{\bibfnamefont{T.~M.~C.} \bibnamefont{Abbott}}
  \bibnamefont{et~al.} (\bibinfo{collaboration}{DES}), \bibinfo{journal}{Phys.
  Rev. D} \textbf{\bibinfo{volume}{98}}, \bibinfo{pages}{043526}
  (\bibinfo{year}{2018}), \eprint{1708.01530}.

\bibitem[{\citenamefont{Abbott et~al.}(2021)}]{DES:2021wwk}
\bibinfo{author}{\bibfnamefont{T.~M.~C.} \bibnamefont{Abbott}}
  \bibnamefont{et~al.} (\bibinfo{collaboration}{DES}) (\bibinfo{year}{2021}),
  \eprint{2105.13549}.

\bibitem[{\citenamefont{Asgari et~al.}(2021)}]{KiDS:2020suj}
\bibinfo{author}{\bibfnamefont{M.}~\bibnamefont{Asgari}} \bibnamefont{et~al.}
  (\bibinfo{collaboration}{KiDS}), \bibinfo{journal}{Astron. Astrophys.}
  \textbf{\bibinfo{volume}{645}}, \bibinfo{pages}{A104} (\bibinfo{year}{2021}),
  \eprint{2007.15633}.

\bibitem[{\citenamefont{Heymans et~al.}(2021)}]{Heymans:2020gsg}
\bibinfo{author}{\bibfnamefont{C.}~\bibnamefont{Heymans}} \bibnamefont{et~al.},
  \bibinfo{journal}{Astron. Astrophys.} \textbf{\bibinfo{volume}{646}},
  \bibinfo{pages}{A140} (\bibinfo{year}{2021}), \eprint{2007.15632}.

\bibitem[{\citenamefont{Nunes and Vagnozzi}(2021)}]{Nunes:2021ipq}
\bibinfo{author}{\bibfnamefont{R.~C.} \bibnamefont{Nunes}} \bibnamefont{and}
  \bibinfo{author}{\bibfnamefont{S.}~\bibnamefont{Vagnozzi}},
  \bibinfo{journal}{Mon. Not. Roy. Astron. Soc.}
  \textbf{\bibinfo{volume}{505}}, \bibinfo{pages}{5427} (\bibinfo{year}{2021}),
  \eprint{2106.01208}.

\bibitem[{\citenamefont{Poulin et~al.}(2019)\citenamefont{Poulin, Smith,
  Karwal, and Kamionkowski}}]{Poulin:2018cxd}
\bibinfo{author}{\bibfnamefont{V.}~\bibnamefont{Poulin}},
  \bibinfo{author}{\bibfnamefont{T.~L.} \bibnamefont{Smith}},
  \bibinfo{author}{\bibfnamefont{T.}~\bibnamefont{Karwal}}, \bibnamefont{and}
  \bibinfo{author}{\bibfnamefont{M.}~\bibnamefont{Kamionkowski}},
  \bibinfo{journal}{Phys. Rev. Lett.} \textbf{\bibinfo{volume}{122}},
  \bibinfo{pages}{221301} (\bibinfo{year}{2019}), \eprint{1811.04083}.

\bibitem[{\citenamefont{Smith et~al.}(2020)\citenamefont{Smith, Poulin, and
  Amin}}]{Smith:2019ihp}
\bibinfo{author}{\bibfnamefont{T.~L.} \bibnamefont{Smith}},
  \bibinfo{author}{\bibfnamefont{V.}~\bibnamefont{Poulin}}, \bibnamefont{and}
  \bibinfo{author}{\bibfnamefont{M.~A.} \bibnamefont{Amin}},
  \bibinfo{journal}{Phys. Rev. D} \textbf{\bibinfo{volume}{101}},
  \bibinfo{pages}{063523} (\bibinfo{year}{2020}), \eprint{1908.06995}.

\bibitem[{\citenamefont{Alcaniz et~al.}(2021)\citenamefont{Alcaniz, Bernal,
  Masiero, and Queiroz}}]{Alcaniz:2019kah}
\bibinfo{author}{\bibfnamefont{J.}~\bibnamefont{Alcaniz}},
  \bibinfo{author}{\bibfnamefont{N.}~\bibnamefont{Bernal}},
  \bibinfo{author}{\bibfnamefont{A.}~\bibnamefont{Masiero}}, \bibnamefont{and}
  \bibinfo{author}{\bibfnamefont{F.~S.} \bibnamefont{Queiroz}},
  \bibinfo{journal}{Phys. Lett. B} \textbf{\bibinfo{volume}{812}},
  \bibinfo{pages}{136008} (\bibinfo{year}{2021}), \eprint{1912.05563}.

\bibitem[{\citenamefont{Zumalacarregui}(2020)}]{Zumalacarregui:2020cjh}
\bibinfo{author}{\bibfnamefont{M.}~\bibnamefont{Zumalacarregui}},
  \bibinfo{journal}{Phys. Rev. D} \textbf{\bibinfo{volume}{102}},
  \bibinfo{pages}{023523} (\bibinfo{year}{2020}), \eprint{2003.06396}.

\bibitem[{\citenamefont{G\'omez-Valent
  et~al.}(2020)\citenamefont{G\'omez-Valent, Pettorino, and
  Amendola}}]{Gomez-Valent:2020mqn}
\bibinfo{author}{\bibfnamefont{A.}~\bibnamefont{G\'omez-Valent}},
  \bibinfo{author}{\bibfnamefont{V.}~\bibnamefont{Pettorino}},
  \bibnamefont{and} \bibinfo{author}{\bibfnamefont{L.}~\bibnamefont{Amendola}},
  \bibinfo{journal}{Phys. Rev. D} \textbf{\bibinfo{volume}{101}},
  \bibinfo{pages}{123513} (\bibinfo{year}{2020}), \eprint{2004.00610}.

\bibitem[{\citenamefont{Ballesteros et~al.}(2020)\citenamefont{Ballesteros,
  Notari, and Rompineve}}]{Ballesteros:2020sik}
\bibinfo{author}{\bibfnamefont{G.}~\bibnamefont{Ballesteros}},
  \bibinfo{author}{\bibfnamefont{A.}~\bibnamefont{Notari}}, \bibnamefont{and}
  \bibinfo{author}{\bibfnamefont{F.}~\bibnamefont{Rompineve}},
  \bibinfo{journal}{JCAP} \textbf{\bibinfo{volume}{11}}, \bibinfo{pages}{024}
  (\bibinfo{year}{2020}), \eprint{2004.05049}.

\bibitem[{\citenamefont{Jim\'enez et~al.}(2021)\citenamefont{Jim\'enez,
  Bettoni, and Brax}}]{Jimenez:2020bgw}
\bibinfo{author}{\bibfnamefont{J.~B.} \bibnamefont{Jim\'enez}},
  \bibinfo{author}{\bibfnamefont{D.}~\bibnamefont{Bettoni}}, \bibnamefont{and}
  \bibinfo{author}{\bibfnamefont{P.}~\bibnamefont{Brax}},
  \bibinfo{journal}{Phys. Rev. D} \textbf{\bibinfo{volume}{103}},
  \bibinfo{pages}{103505} (\bibinfo{year}{2021}), \eprint{2004.13677}.

\bibitem[{\citenamefont{Di~Valentino
  et~al.}(2021{\natexlab{a}})\citenamefont{Di~Valentino, Mukherjee, and
  Sen}}]{DiValentino:2020naf}
\bibinfo{author}{\bibfnamefont{E.}~\bibnamefont{Di~Valentino}},
  \bibinfo{author}{\bibfnamefont{A.}~\bibnamefont{Mukherjee}},
  \bibnamefont{and} \bibinfo{author}{\bibfnamefont{A.~A.} \bibnamefont{Sen}},
  \bibinfo{journal}{Entropy} \textbf{\bibinfo{volume}{23}},
  \bibinfo{pages}{404} (\bibinfo{year}{2021}{\natexlab{a}}),
  \eprint{2005.12587}.

\bibitem[{\citenamefont{Banerjee et~al.}(2021)\citenamefont{Banerjee, Cai,
  Heisenberg, Colg\'ain, Sheikh-Jabbari, and Yang}}]{Banerjee:2020xcn}
\bibinfo{author}{\bibfnamefont{A.}~\bibnamefont{Banerjee}},
  \bibinfo{author}{\bibfnamefont{H.}~\bibnamefont{Cai}},
  \bibinfo{author}{\bibfnamefont{L.}~\bibnamefont{Heisenberg}},
  \bibinfo{author}{\bibfnamefont{E.~O.} \bibnamefont{Colg\'ain}},
  \bibinfo{author}{\bibfnamefont{M.~M.} \bibnamefont{Sheikh-Jabbari}},
  \bibnamefont{and} \bibinfo{author}{\bibfnamefont{T.}~\bibnamefont{Yang}},
  \bibinfo{journal}{Phys. Rev. D} \textbf{\bibinfo{volume}{103}},
  \bibinfo{pages}{L081305} (\bibinfo{year}{2021}), \eprint{2006.00244}.

\bibitem[{\citenamefont{Krishnan et~al.}(2021)\citenamefont{Krishnan, Mohayaee,
  Colg\'ain, Sheikh-Jabbari, and Yin}}]{Krishnan:2021dyb}
\bibinfo{author}{\bibfnamefont{C.}~\bibnamefont{Krishnan}},
  \bibinfo{author}{\bibfnamefont{R.}~\bibnamefont{Mohayaee}},
  \bibinfo{author}{\bibfnamefont{E.~O.} \bibnamefont{Colg\'ain}},
  \bibinfo{author}{\bibfnamefont{M.~M.} \bibnamefont{Sheikh-Jabbari}},
  \bibnamefont{and} \bibinfo{author}{\bibfnamefont{L.}~\bibnamefont{Yin}},
  \bibinfo{journal}{Class. Quant. Grav.} \textbf{\bibinfo{volume}{38}},
  \bibinfo{pages}{184001} (\bibinfo{year}{2021}), \eprint{2105.09790}.

\bibitem[{\citenamefont{Teng et~al.}(2021)\citenamefont{Teng, Lee, and
  Ng}}]{Teng:2021cvy}
\bibinfo{author}{\bibfnamefont{Y.-P.} \bibnamefont{Teng}},
  \bibinfo{author}{\bibfnamefont{W.}~\bibnamefont{Lee}}, \bibnamefont{and}
  \bibinfo{author}{\bibfnamefont{K.-W.} \bibnamefont{Ng}},
  \bibinfo{journal}{Phys. Rev. D} \textbf{\bibinfo{volume}{104}},
  \bibinfo{pages}{083519} (\bibinfo{year}{2021}), \eprint{2105.02667}.

\bibitem[{\citenamefont{Ballardini et~al.}(2021)\citenamefont{Ballardini,
  Finelli, and Sapone}}]{Ballardini:2021evv}
\bibinfo{author}{\bibfnamefont{M.}~\bibnamefont{Ballardini}},
  \bibinfo{author}{\bibfnamefont{F.}~\bibnamefont{Finelli}}, \bibnamefont{and}
  \bibinfo{author}{\bibfnamefont{D.}~\bibnamefont{Sapone}}
  (\bibinfo{year}{2021}), \eprint{2111.09168}.

\bibitem[{\citenamefont{Braglia et~al.}(2021)\citenamefont{Braglia, Ballardini,
  Finelli, and Koyama}}]{Braglia:2020auw}
\bibinfo{author}{\bibfnamefont{M.}~\bibnamefont{Braglia}},
  \bibinfo{author}{\bibfnamefont{M.}~\bibnamefont{Ballardini}},
  \bibinfo{author}{\bibfnamefont{F.}~\bibnamefont{Finelli}}, \bibnamefont{and}
  \bibinfo{author}{\bibfnamefont{K.}~\bibnamefont{Koyama}},
  \bibinfo{journal}{Phys. Rev. D} \textbf{\bibinfo{volume}{103}},
  \bibinfo{pages}{043528} (\bibinfo{year}{2021}), \eprint{2011.12934}.

\bibitem[{\citenamefont{Braglia et~al.}(2020)\citenamefont{Braglia, Ballardini,
  Emond, Finelli, Gumrukcuoglu, Koyama, and Paoletti}}]{Braglia:2020iik}
\bibinfo{author}{\bibfnamefont{M.}~\bibnamefont{Braglia}},
  \bibinfo{author}{\bibfnamefont{M.}~\bibnamefont{Ballardini}},
  \bibinfo{author}{\bibfnamefont{W.~T.} \bibnamefont{Emond}},
  \bibinfo{author}{\bibfnamefont{F.}~\bibnamefont{Finelli}},
  \bibinfo{author}{\bibfnamefont{A.~E.} \bibnamefont{Gumrukcuoglu}},
  \bibinfo{author}{\bibfnamefont{K.}~\bibnamefont{Koyama}}, \bibnamefont{and}
  \bibinfo{author}{\bibfnamefont{D.}~\bibnamefont{Paoletti}},
  \bibinfo{journal}{Phys. Rev. D} \textbf{\bibinfo{volume}{102}},
  \bibinfo{pages}{023529} (\bibinfo{year}{2020}), \eprint{2004.11161}.

\bibitem[{\citenamefont{Lambiase et~al.}(2019)\citenamefont{Lambiase, Mohanty,
  Narang, and Parashari}}]{Lambiase:2018ows}
\bibinfo{author}{\bibfnamefont{G.}~\bibnamefont{Lambiase}},
  \bibinfo{author}{\bibfnamefont{S.}~\bibnamefont{Mohanty}},
  \bibinfo{author}{\bibfnamefont{A.}~\bibnamefont{Narang}}, \bibnamefont{and}
  \bibinfo{author}{\bibfnamefont{P.}~\bibnamefont{Parashari}},
  \bibinfo{journal}{Eur. Phys. J. C} \textbf{\bibinfo{volume}{79}},
  \bibinfo{pages}{141} (\bibinfo{year}{2019}), \eprint{1804.07154}.

\bibitem[{\citenamefont{Keeley et~al.}(2019)\citenamefont{Keeley, Joudaki,
  Kaplinghat, and Kirkby}}]{Keeley:2019esp}
\bibinfo{author}{\bibfnamefont{R.~E.} \bibnamefont{Keeley}},
  \bibinfo{author}{\bibfnamefont{S.}~\bibnamefont{Joudaki}},
  \bibinfo{author}{\bibfnamefont{M.}~\bibnamefont{Kaplinghat}},
  \bibnamefont{and} \bibinfo{author}{\bibfnamefont{D.}~\bibnamefont{Kirkby}},
  \bibinfo{journal}{JCAP} \textbf{\bibinfo{volume}{12}}, \bibinfo{pages}{035}
  (\bibinfo{year}{2019}), \eprint{1905.10198}.

\bibitem[{\citenamefont{Di~Valentino et~al.}(2020)\citenamefont{Di~Valentino,
  Melchiorri, Mena, and Vagnozzi}}]{DiValentino:2019ffd}
\bibinfo{author}{\bibfnamefont{E.}~\bibnamefont{Di~Valentino}},
  \bibinfo{author}{\bibfnamefont{A.}~\bibnamefont{Melchiorri}},
  \bibinfo{author}{\bibfnamefont{O.}~\bibnamefont{Mena}}, \bibnamefont{and}
  \bibinfo{author}{\bibfnamefont{S.}~\bibnamefont{Vagnozzi}},
  \bibinfo{journal}{Phys. Dark Univ.} \textbf{\bibinfo{volume}{30}},
  \bibinfo{pages}{100666} (\bibinfo{year}{2020}), \eprint{1908.04281}.

\bibitem[{\citenamefont{Jedamzik et~al.}(2021)\citenamefont{Jedamzik, Pogosian,
  and Zhao}}]{Jedamzik:2020zmd}
\bibinfo{author}{\bibfnamefont{K.}~\bibnamefont{Jedamzik}},
  \bibinfo{author}{\bibfnamefont{L.}~\bibnamefont{Pogosian}}, \bibnamefont{and}
  \bibinfo{author}{\bibfnamefont{G.-B.} \bibnamefont{Zhao}},
  \bibinfo{journal}{Commun. in Phys.} \textbf{\bibinfo{volume}{4}},
  \bibinfo{pages}{123} (\bibinfo{year}{2021}), \eprint{2010.04158}.

\bibitem[{\citenamefont{Clark et~al.}(2021)\citenamefont{Clark, Vattis, Fan,
  and Koushiappas}}]{Clark:2021hlo}
\bibinfo{author}{\bibfnamefont{S.~J.} \bibnamefont{Clark}},
  \bibinfo{author}{\bibfnamefont{K.}~\bibnamefont{Vattis}},
  \bibinfo{author}{\bibfnamefont{J.}~\bibnamefont{Fan}}, \bibnamefont{and}
  \bibinfo{author}{\bibfnamefont{S.~M.} \bibnamefont{Koushiappas}}
  (\bibinfo{year}{2021}), \eprint{2110.09562}.

\bibitem[{\citenamefont{Sol\`a~Peracaula
  et~al.}(2021)\citenamefont{Sol\`a~Peracaula, G\'omez-Valent, de~Cruz~Perez,
  and Moreno-Pulido}}]{SolaPeracaula:2021gxi}
\bibinfo{author}{\bibfnamefont{J.}~\bibnamefont{Sol\`a~Peracaula}},
  \bibinfo{author}{\bibfnamefont{A.}~\bibnamefont{G\'omez-Valent}},
  \bibinfo{author}{\bibfnamefont{J.}~\bibnamefont{de~Cruz~Perez}},
  \bibnamefont{and}
  \bibinfo{author}{\bibfnamefont{C.}~\bibnamefont{Moreno-Pulido}},
  \bibinfo{journal}{EPL} \textbf{\bibinfo{volume}{134}}, \bibinfo{pages}{19001}
  (\bibinfo{year}{2021}), \eprint{2102.12758}.

\bibitem[{\citenamefont{Alestas and Perivolaropoulos}(2021)}]{Alestas:2021xes}
\bibinfo{author}{\bibfnamefont{G.}~\bibnamefont{Alestas}} \bibnamefont{and}
  \bibinfo{author}{\bibfnamefont{L.}~\bibnamefont{Perivolaropoulos}},
  \bibinfo{journal}{Mon. Not. Roy. Astron. Soc.}
  \textbf{\bibinfo{volume}{504}}, \bibinfo{pages}{3956} (\bibinfo{year}{2021}),
  \eprint{2103.04045}.

\bibitem[{\citenamefont{Sch\"oneberg et~al.}(2021)\citenamefont{Sch\"oneberg,
  Franco~Abell\'an, P\'erez~S\'anchez, Witte, Poulin, and
  Lesgourgues}}]{Schoneberg:2021qvd}
\bibinfo{author}{\bibfnamefont{N.}~\bibnamefont{Sch\"oneberg}},
  \bibinfo{author}{\bibfnamefont{G.}~\bibnamefont{Franco~Abell\'an}},
  \bibinfo{author}{\bibfnamefont{A.}~\bibnamefont{P\'erez~S\'anchez}},
  \bibinfo{author}{\bibfnamefont{S.~J.} \bibnamefont{Witte}},
  \bibinfo{author}{\bibfnamefont{V.}~\bibnamefont{Poulin}}, \bibnamefont{and}
  \bibinfo{author}{\bibfnamefont{J.}~\bibnamefont{Lesgourgues}}
  (\bibinfo{year}{2021}), \eprint{2107.10291}.

\bibitem[{\citenamefont{Alestas et~al.}(2021)\citenamefont{Alestas, Camarena,
  Di~Valentino, Kazantzidis, Marra, Nesseris, and
  Perivolaropoulos}}]{Alestas:2021luu}
\bibinfo{author}{\bibfnamefont{G.}~\bibnamefont{Alestas}},
  \bibinfo{author}{\bibfnamefont{D.}~\bibnamefont{Camarena}},
  \bibinfo{author}{\bibfnamefont{E.}~\bibnamefont{Di~Valentino}},
  \bibinfo{author}{\bibfnamefont{L.}~\bibnamefont{Kazantzidis}},
  \bibinfo{author}{\bibfnamefont{V.}~\bibnamefont{Marra}},
  \bibinfo{author}{\bibfnamefont{S.}~\bibnamefont{Nesseris}}, \bibnamefont{and}
  \bibinfo{author}{\bibfnamefont{L.}~\bibnamefont{Perivolaropoulos}}
  (\bibinfo{year}{2021}), \eprint{2110.04336}.

\bibitem[{\citenamefont{Ye et~al.}(2021)\citenamefont{Ye, Zhang, and
  Piao}}]{Ye:2021iwa}
\bibinfo{author}{\bibfnamefont{G.}~\bibnamefont{Ye}},
  \bibinfo{author}{\bibfnamefont{J.}~\bibnamefont{Zhang}}, \bibnamefont{and}
  \bibinfo{author}{\bibfnamefont{Y.-S.} \bibnamefont{Piao}}
  (\bibinfo{year}{2021}), \eprint{2107.13391}.

\bibitem[{\citenamefont{Riess}(2019)}]{Riess:2019qba}
\bibinfo{author}{\bibfnamefont{A.~G.} \bibnamefont{Riess}},
  \bibinfo{journal}{Nature Rev. Phys.} \textbf{\bibinfo{volume}{2}},
  \bibinfo{pages}{10} (\bibinfo{year}{2019}), \eprint{2001.03624}.

\bibitem[{\citenamefont{Knox and Millea}(2020)}]{Knox:2019rjx}
\bibinfo{author}{\bibfnamefont{L.}~\bibnamefont{Knox}} \bibnamefont{and}
  \bibinfo{author}{\bibfnamefont{M.}~\bibnamefont{Millea}},
  \bibinfo{journal}{Phys. Rev. D} \textbf{\bibinfo{volume}{101}},
  \bibinfo{pages}{043533} (\bibinfo{year}{2020}), \eprint{1908.03663}.

\bibitem[{\citenamefont{Di~Valentino
  et~al.}(2021{\natexlab{b}})}]{DiValentino:2020vvd}
\bibinfo{author}{\bibfnamefont{E.}~\bibnamefont{Di~Valentino}}
  \bibnamefont{et~al.}, \bibinfo{journal}{Astropart. Phys.}
  \textbf{\bibinfo{volume}{131}}, \bibinfo{pages}{102604}
  (\bibinfo{year}{2021}{\natexlab{b}}), \eprint{2008.11285}.

\bibitem[{\citenamefont{Di~Valentino
  et~al.}(2021{\natexlab{c}})}]{DiValentino:2020zio}
\bibinfo{author}{\bibfnamefont{E.}~\bibnamefont{Di~Valentino}}
  \bibnamefont{et~al.}, \bibinfo{journal}{Astropart. Phys.}
  \textbf{\bibinfo{volume}{131}}, \bibinfo{pages}{102605}
  (\bibinfo{year}{2021}{\natexlab{c}}), \eprint{2008.11284}.

\bibitem[{\citenamefont{Di~Valentino
  et~al.}(2021{\natexlab{d}})\citenamefont{Di~Valentino, Mena, Pan, Visinelli,
  Yang, Melchiorri, Mota, Riess, and Silk}}]{DiValentino:2021izs}
\bibinfo{author}{\bibfnamefont{E.}~\bibnamefont{Di~Valentino}},
  \bibinfo{author}{\bibfnamefont{O.}~\bibnamefont{Mena}},
  \bibinfo{author}{\bibfnamefont{S.}~\bibnamefont{Pan}},
  \bibinfo{author}{\bibfnamefont{L.}~\bibnamefont{Visinelli}},
  \bibinfo{author}{\bibfnamefont{W.}~\bibnamefont{Yang}},
  \bibinfo{author}{\bibfnamefont{A.}~\bibnamefont{Melchiorri}},
  \bibinfo{author}{\bibfnamefont{D.~F.} \bibnamefont{Mota}},
  \bibinfo{author}{\bibfnamefont{A.~G.} \bibnamefont{Riess}}, \bibnamefont{and}
  \bibinfo{author}{\bibfnamefont{J.}~\bibnamefont{Silk}},
  \bibinfo{journal}{Class. Quant. Grav.} \textbf{\bibinfo{volume}{38}},
  \bibinfo{pages}{153001} (\bibinfo{year}{2021}{\natexlab{d}}),
  \eprint{2103.01183}.

\bibitem[{\citenamefont{Perivolaropoulos and
  Skara}(2021)}]{Perivolaropoulos:2021jda}
\bibinfo{author}{\bibfnamefont{L.}~\bibnamefont{Perivolaropoulos}}
  \bibnamefont{and} \bibinfo{author}{\bibfnamefont{F.}~\bibnamefont{Skara}}
  (\bibinfo{year}{2021}), \eprint{2105.05208}.

\bibitem[{\citenamefont{Renk et~al.}(2017)\citenamefont{Renk, Zumalac\'arregui,
  Montanari, and Barreira}}]{Renk:2017rzu}
\bibinfo{author}{\bibfnamefont{J.}~\bibnamefont{Renk}},
  \bibinfo{author}{\bibfnamefont{M.}~\bibnamefont{Zumalac\'arregui}},
  \bibinfo{author}{\bibfnamefont{F.}~\bibnamefont{Montanari}},
  \bibnamefont{and} \bibinfo{author}{\bibfnamefont{A.}~\bibnamefont{Barreira}},
  \bibinfo{journal}{JCAP} \textbf{\bibinfo{volume}{10}}, \bibinfo{pages}{020}
  (\bibinfo{year}{2017}), \eprint{1707.02263}.

\bibitem[{\citenamefont{Frusciante et~al.}(2020)\citenamefont{Frusciante,
  Peirone, Atayde, and De~Felice}}]{Frusciante:2019puu}
\bibinfo{author}{\bibfnamefont{N.}~\bibnamefont{Frusciante}},
  \bibinfo{author}{\bibfnamefont{S.}~\bibnamefont{Peirone}},
  \bibinfo{author}{\bibfnamefont{L.}~\bibnamefont{Atayde}}, \bibnamefont{and}
  \bibinfo{author}{\bibfnamefont{A.}~\bibnamefont{De~Felice}},
  \bibinfo{journal}{Phys. Rev. D} \textbf{\bibinfo{volume}{101}},
  \bibinfo{pages}{064001} (\bibinfo{year}{2020}), \eprint{1912.07586}.

\bibitem[{\citenamefont{de~Felice et~al.}(2017)\citenamefont{de~Felice,
  Heisenberg, and Tsujikawa}}]{deFelice:2017paw}
\bibinfo{author}{\bibfnamefont{A.}~\bibnamefont{de~Felice}},
  \bibinfo{author}{\bibfnamefont{L.}~\bibnamefont{Heisenberg}},
  \bibnamefont{and}
  \bibinfo{author}{\bibfnamefont{S.}~\bibnamefont{Tsujikawa}},
  \bibinfo{journal}{Phys. Rev. D} \textbf{\bibinfo{volume}{95}},
  \bibinfo{pages}{123540} (\bibinfo{year}{2017}), \eprint{1703.09573}.

\bibitem[{\citenamefont{De~Felice et~al.}(2020)\citenamefont{De~Felice, Geng,
  Pookkillath, and Yin}}]{DeFelice:2020sdq}
\bibinfo{author}{\bibfnamefont{A.}~\bibnamefont{De~Felice}},
  \bibinfo{author}{\bibfnamefont{C.-Q.} \bibnamefont{Geng}},
  \bibinfo{author}{\bibfnamefont{M.~C.} \bibnamefont{Pookkillath}},
  \bibnamefont{and} \bibinfo{author}{\bibfnamefont{L.}~\bibnamefont{Yin}},
  \bibinfo{journal}{JCAP} \textbf{\bibinfo{volume}{08}}, \bibinfo{pages}{038}
  (\bibinfo{year}{2020}), \eprint{2002.06782}.

\bibitem[{\citenamefont{Heisenberg and
  Villarrubia-Rojo}(2021)}]{Heisenberg:2020xak}
\bibinfo{author}{\bibfnamefont{L.}~\bibnamefont{Heisenberg}} \bibnamefont{and}
  \bibinfo{author}{\bibfnamefont{H.}~\bibnamefont{Villarrubia-Rojo}},
  \bibinfo{journal}{JCAP} \textbf{\bibinfo{volume}{03}}, \bibinfo{pages}{032}
  (\bibinfo{year}{2021}), \eprint{2010.00513}.

\bibitem[{\citenamefont{Chen et~al.}(2019)\citenamefont{Chen, Huang, and
  Wang}}]{Chen:2018dbv}
\bibinfo{author}{\bibfnamefont{L.}~\bibnamefont{Chen}},
  \bibinfo{author}{\bibfnamefont{Q.-G.} \bibnamefont{Huang}}, \bibnamefont{and}
  \bibinfo{author}{\bibfnamefont{K.}~\bibnamefont{Wang}},
  \bibinfo{journal}{JCAP} \textbf{\bibinfo{volume}{02}}, \bibinfo{pages}{028}
  (\bibinfo{year}{2019}), \eprint{1808.05724}.

\bibitem[{\citenamefont{Dodelson and Schmidt}(2020)}]{dodelson2020modern}
\bibinfo{author}{\bibfnamefont{S.}~\bibnamefont{Dodelson}} \bibnamefont{and}
  \bibinfo{author}{\bibfnamefont{F.}~\bibnamefont{Schmidt}},
  \emph{\bibinfo{title}{Modern Cosmology}} (\bibinfo{publisher}{Elsevier
  Science}, \bibinfo{year}{2020}), ISBN \bibinfo{isbn}{9780128159484}.

\bibitem[{\citenamefont{Eisenstein and Hu}(1998)}]{Eisenstein:1997ik}
\bibinfo{author}{\bibfnamefont{D.~J.} \bibnamefont{Eisenstein}}
  \bibnamefont{and} \bibinfo{author}{\bibfnamefont{W.}~\bibnamefont{Hu}},
  \bibinfo{journal}{Astrophys. J.} \textbf{\bibinfo{volume}{496}},
  \bibinfo{pages}{605} (\bibinfo{year}{1998}), \eprint{astro-ph/9709112}.

\bibitem[{\citenamefont{Camarena and Marra}(2021)}]{Camarena:2021jlr}
\bibinfo{author}{\bibfnamefont{D.}~\bibnamefont{Camarena}} \bibnamefont{and}
  \bibinfo{author}{\bibfnamefont{V.}~\bibnamefont{Marra}},
  \bibinfo{journal}{Mon. Not. Roy. Astron. Soc.}
  \textbf{\bibinfo{volume}{504}}, \bibinfo{pages}{5164} (\bibinfo{year}{2021}),
  \eprint{2101.08641}.

\bibitem[{\citenamefont{Scolnic et~al.}(2018)}]{Pan-STARRS1:2017jku}
\bibinfo{author}{\bibfnamefont{D.~M.} \bibnamefont{Scolnic}}
  \bibnamefont{et~al.} (\bibinfo{collaboration}{Pan-STARRS1}),
  \bibinfo{journal}{Astrophys. J.} \textbf{\bibinfo{volume}{859}},
  \bibinfo{pages}{101} (\bibinfo{year}{2018}), \eprint{1710.00845}.

\bibitem[{\citenamefont{Benevento et~al.}(2020)\citenamefont{Benevento, Hu, and
  Raveri}}]{Benevento:2020fev}
\bibinfo{author}{\bibfnamefont{G.}~\bibnamefont{Benevento}},
  \bibinfo{author}{\bibfnamefont{W.}~\bibnamefont{Hu}}, \bibnamefont{and}
  \bibinfo{author}{\bibfnamefont{M.}~\bibnamefont{Raveri}},
  \bibinfo{journal}{Phys. Rev. D} \textbf{\bibinfo{volume}{101}},
  \bibinfo{pages}{103517} (\bibinfo{year}{2020}), \eprint{2002.11707}.

\bibitem[{\citenamefont{Chevallier and Polarski}(2001)}]{Chevallier:2000qy}
\bibinfo{author}{\bibfnamefont{M.}~\bibnamefont{Chevallier}} \bibnamefont{and}
  \bibinfo{author}{\bibfnamefont{D.}~\bibnamefont{Polarski}},
  \bibinfo{journal}{Int. J. Mod. Phys. D} \textbf{\bibinfo{volume}{10}},
  \bibinfo{pages}{213} (\bibinfo{year}{2001}), \eprint{gr-qc/0009008}.

\bibitem[{\citenamefont{Linder}(2003)}]{Linder:2002et}
\bibinfo{author}{\bibfnamefont{E.~V.} \bibnamefont{Linder}},
  \bibinfo{journal}{Phys. Rev. Lett.} \textbf{\bibinfo{volume}{90}},
  \bibinfo{pages}{091301} (\bibinfo{year}{2003}), \eprint{astro-ph/0208512}.

\bibitem[{\citenamefont{Blas et~al.}(2011)\citenamefont{Blas, Lesgourgues, and
  Tram}}]{Blas:2011rf}
\bibinfo{author}{\bibfnamefont{D.}~\bibnamefont{Blas}},
  \bibinfo{author}{\bibfnamefont{J.}~\bibnamefont{Lesgourgues}},
  \bibnamefont{and} \bibinfo{author}{\bibfnamefont{T.}~\bibnamefont{Tram}},
  \bibinfo{journal}{JCAP} \textbf{\bibinfo{volume}{07}}, \bibinfo{pages}{034}
  (\bibinfo{year}{2011}), \eprint{1104.2933}.

\end{thebibliography}

\end{document}